\documentclass[twocolumn]{aastex7}

\usepackage{CJK}
\usepackage[normalem]{ulem}

\usepackage{threeparttable} 

\usepackage{float}
\makeatletter
\let\newfloat\newfloat@ltx
\makeatother
\usepackage{algorithm}
\usepackage{algorithmicx}
\usepackage{algpseudocode}
\usepackage{amsmath}
\usepackage{subfigure}
\usepackage{longtable,booktabs}
\usepackage{tablefootnote} 
\usepackage{threeparttable}
\usepackage{appendix}
\usepackage{multirow}
\usepackage{amsmath,amssymb,amsfonts}
\graphicspath{{./}{figures/}}
\usepackage{xcolor}
\usepackage{soul}
\usepackage{hyperref}

\newcommand{\nt}{N_{\mathrm{t}}}
\newcommand{\rt}{r_{\mathrm{t}}}
\newcommand{\rh}{r_{\mathrm{h}}}
\newcommand{\koneave}{\overline{k_1}}

\newcommand{\tbave}{\overline{t_{\mathrm{b}}}}
\newcommand{\koneq}{k_1(q)}
\newcommand{\tb}{t_{\mathrm{b}}}
\newcommand{\tbq}{t_{\mathrm{b}}(q)}
\newcommand{\trh}{t_{\mathrm{rh}}}

\newcommand{\tch}{t_{\mathrm{ch}}}

\newcommand{\tms}{t_{\mathrm{ms}}}
\newcommand{\tmsb}{t_{\mathrm{msb}}}

\newcommand{\tobs}{t_{\mathrm{obs}}}

\begin{document}
\begin{CJK*}{UTF8}{gbsn}

\defcitealias{47-Hunt.2023}{\textsc{H23}}
\defcitealias{64-Marks.2012}{\textsc{M12}}

\title{Binary disruption during the early phase of open clusters }

\author[0009-0007-2280-1254]{Zepeng Zheng}
\affiliation{School of Physics and Astronomy, Sun Yat-sen University, Daxue Road, Zhuhai, 519082, China}
\email[]{ZP.Z.:zhengzp3@mail2.sysu.edu.cn}  

\author[0000-0001-8713-0366]{Long Wang}
\email[show]{L.W.: wanglong8@sysu.edu.cn}
\affiliation{School of Physics and Astronomy, Sun Yat-sen University, Daxue Road, Zhuhai, 519082, China}
\affiliation{CSST Science Center for the Guangdong-Hong Kong-Macau Greater Bay Area, Zhuhai, 519082, China}

\author[0000-0001-8713-0366]{Holger Baumgardt}
\affiliation{School of Mathematics and Physics, The University of Queensland, St. Lucia, Qld 4072, Australia}
\email{}  

\begin{abstract}

The binary fraction in young open clusters exceeds that of field stars, making the study of binary dynamical evolution in clusters essential for understanding the origins and evolution of field binaries. Using N-body simulations based on Gaia DR3 open cluster observations and assuming a 100\% primordial binary fraction, we investigated the early evolution of binary survival fractions in open clusters. We find that binary disruption has two stages, an initial rapid decline followed by a slower decrease, well described by two piecewise linear functions. The early disruption rate, $k_1$, follows a power-law relation with the cluster's initial density ($ \rho_\mathrm{0} $), with an index of approximately 0.56, driven by the disruption of wide binaries via close encounters.
The transition time between the two phases, $t_\mathrm{b}$, also exhibits a power-law dependence on $\rho_\mathrm{0}$ with an index of about -0.46.
The disruption rate also depends on binary parameters: high-$q$ and wide binaries are disrupted faster, while the dependence on eccentricity $e$ is less clear, likely due to its strong evolution.
We developed and publicly released a Python tool to predict binary survival fraction evolution based on $\rho_0$, $P$ and $q$. 
Additionally, we also investigate how open cluster binaries contribute to the field population, and find that the escaped stars have a systematically lower binary fraction, likely due to mass segregation. 
Both populations show similar distributions of $ P $ and $e$, but lower-$q$ systems preferentially remain bound within clusters, the origin of which remains uncertain.

\end{abstract}


\section{Introduction}

Studies show that stars form in giant molecular clouds and are often assembled into star clusters \citep{3-Lada.2003,35-Lada.2010,36-Zwart.2010}. Over the past few decades, a large number of observations have shown that the fraction of binary and higher-order systems in star-forming regions as well as in young open clusters is very high. For example, \cite{26-Duchene.1998} conducted a comprehensive study of low-mass binary stars in several star-forming regions and found a significant binary excess among Taurus members. Under the assumption that the period distribution of these stars is consistent with that of main-sequence binaries, up to 95\% of stars are found in multiple star systems. \cite{27-Connelley.2008} investigated the evolution of binary separation distributions for Class I Young Stellar Objects (YSOs) in star-forming regions such as Taurus, Ophiuchus and Orion, and concluded that nearly all stars form in binary or multiple groups. During the Class 0 phase, the earliest, deeply embedded stage, many objects are part of non-hierarchical multiple systems. By the end of the Class I phase, when the protostar is more evolved with a less massive envelope, wide-separation binaries undergo significant disruption, leading to a significant decrease in binary frequency. This evolutionary trend in binary frequency was further validated by \cite{28-Chen2013} through a survey of Class 0 YSOs. Without completeness corrections, the estimated binary fraction for Class 0 YSOs is approximately 64\%, roughly twice that of Class I YSOs, indicating substantial binary disruption during the evolution from Class 0 to Class I YSOs. \cite{9-Sana.2012} found that over 70\% of O-type stars in open clusters experience binary interactions. \cite{10-Donada.2023} studied the binary fraction in 202 Galactic open clusters, finding that for high mass-ratio binary systems ($q > 0.6$), the binary fraction can reach up to 80\%. \cite{1-Ramirez-Tannus.2024} estimated the intrinsic binary fraction of the very young star cluster M17 to be 87\%.

In contrast, field stars exhibit a relatively lower binary proportion. For instance,  \cite{22-Duquennoy.1991} studied 164 solar-type stars in the solar neighborhood and found a multiple star fraction of 43\%. \cite{23-Raghavan.2010} surveyed 454 stars within 25 pc of the Sun in the Hipparcos catalog \citep{82-HIPPARCOS.1997} and found a binary fraction of approximately 33\%. \cite{2-Duchene.2018} conducted a survey of close visual binaries among members of the Orion Nebula Cluster with masses between $0.3 M_\odot$ and $2.0 M_\odot$, finding that their companion star fraction is twice that of field stars.

Why do the binary fractions in star-forming regions and young open clusters differ so significantly from those of field stars? Understanding the binary fraction in star clusters and its evolutionary processes is crucial for addressing many important scientific questions.
An important open question concerns the origin of binary systems among field stars. One hypothesis proposes that most field binaries originate from star clusters. 
Using N-body simulations, \cite{29-Kroupa.1995} showed that open clusters with a 100\% primordial binary fraction undergo rapid dynamical disruption, with the surviving binaries after cluster dissolution reproducing the observed properties of field binaries as found by \cite{22-Duquennoy.1991,39-Simon.1992,40-Leinert.1993,41-Richichi.1994}.
Later observations, such as \cite{23-Raghavan.2010}, further confirmed this picture.
Building on this idea, \cite{60-Marks.2011} developed a dynamical operator formalism to predict the binary evolution in clusters, and similar models have  reproduced the binary properties observed in several young star-forming regions, such as Taurus, $\rho$ Ophiuchus, and Chamaeleon \citep{64-Marks.2012}.
However, other observations reveal binary fractions that deviate from this model \citep{76-Moe.2017,18-Offner.2023}, 
leaving the formation pathway of field binaries still uncertain. Improved models for dynamical evolution of binary populations within clusters are therefore crucial for addressing this long-standing problem.

In star clusters, 
two main physical mechanisms determine the fate of binaries with different orbital separations: orbital decay \citep{79-Stahler.2010} and dynamical interactions \citep{44-Heggie.1975}.
Orbital decay primarily affects short-period binaries embedded in the gaseous environment of star-forming clusters. These binaries experience gas-induced dynamical friction \citep{81-Korntreff.2012}, gas accretion onto their envelopes, and disk interactions (if present) \citep{80-Bate.2002}. These processes reduce orbital energy and angular momentum, shrinking binary separation,  occasionally resulting in mergers. 
In contrast,  dynamical interactions mainly influence long-period binaries. Close encounters can perturb binary orbits, exchange components, or disrupt binaries \citep{29-Kroupa.1995}. This process depends on the cluster's hard$-$soft boundary, where encounters disrupt soft binaries with low binding energies while hard binaries become more tightly bound \citep{44-Heggie.1975,45-Hills.1975}.
Together, these mechanisms preferentially disrupt short- and long-period binaries, leaving a surviving binary population dominated by intermediate-period systems. 

The evolution of the binary fraction is closely linked to cluster mass and density 
as shown in both observations 
and simulations. 
For example, \citet{8-Sollima.2010} found an inverse correlation between the binary fraction in cluster cores and cluster mass when investigating open clusters in the Milky Way. 
Through a series of $N$-body simulations covering different initial cluster masses and half-mass radii, \citet{60-Marks.2011} showed that star clusters with larger masses and smaller half-mass radii experience stronger binary disruption during their early evolutionary stages. 
Similarly, \citet{37-Krug.2012} performed $N$-body simulations of dense clusters containing different numbers of stars and found that systems with more stars$-$and thus higher initial mass and density$-$exhibit a faster decline in their binary fraction over time.
More recently, \citet{10-Donada.2023} measured the binary fractions of 202 open clusters using Gaia data. Although a clear dependence of the binary fraction on cluster mass was not firmly established, they found that massive clusters tend to have lower binary fractions, whereas low-mass clusters show a wider range of binary fractions, including cases with relatively high binary fraction.

Previous theoretical studies of binary dynamical evolution in open clusters face several limitations. Most rely on simplified analyses or restricted $N$-body simulations (e.g. no stellar evolution, limited primordial binary population and simplified Galactic potentials). These simplifications limit the ability to understand how binary evolution depends on cluster properties and orbital parameters such as period ($P$), mass ratio ($q$) and eccentricity ($e$), as well as how binaries escape to the field. 

Additionally, previous studies suffer from limited observational and theoretical samples of open clusters, which can introduce biases in the study of binary disruption. The recent release of Gaia DR3 \citep{62-Gaia.2023} has provided a wealth of precise observational data for cluster studies. Notably, \cite{47-Hunt.2023} constructed a catalog of 7,167 open clusters within the Milky Way based on Gaia DR3 observations, offering a substantial sample for the study and simulation of open clusters. Therefore, in this work, we will utilize the open cluster catalog published by \cite{47-Hunt.2023} (hereafter referred to as \citetalias{47-Hunt.2023}) to select a subset of valuable observed clusters for extensive N-body numerical simulations, thereby establishing a comprehensive simulation database. Based on this database, we systematically investigate how the binary survival fraction in different open clusters, as observed by Gaia, evolves over time, as well as its dependence on $P$, $q$, and initial density $\rho_\mathrm{0}$, defined as the mean stellar mass density within the initial half-mass radius ($r_\mathrm{h,0}$).
We also explore the differences between binaries that escape clusters after tidal stripping and those that remain within clusters, providing theoretical support to address the inconsistencies between field and cluster binary populations.

In Section \ref{sec:2}, we introduce the N-body simulation program \textsc{petar} used for star cluster simulations, as well as the selection process for the simulated star cluster samples and the setup of initial conditions. In Section \ref{sec:3}, we analyze the evolution of the binary survival fraction in simulated star clusters over time, quantify the binary disruption process using equations, and examine the relationship between the binary survival fraction evolution and binary parameters such as $q$ and $P$. Finally, we compare the binary survival fractions, $P$, $q$, $e$ and mass inside and outside the tidal radius of the clusters. In Section \ref{sec:4}, we discuss the assumptions in our models, limitations and future improvements. 
Section \ref{sec:5} provides a summary of our work.

\section{Methods} \label{sec:2}

\subsection{\textsc{petar} Simulation Code}\label{sec:2.1}
In this study, we employed the N-body simulation code \textsc{petar} \citep{66-Wang.2020} to conduct numerical simulations of open star clusters. As a high-performance N-body simulation code, \textsc{petar} combines the particle-tree particle-particle ($\rm P^3T$) method \citep{67-Oshino.2011} with the slow-down algorithmic regularization method (SDAR) \citep{68-Wang.2020}, enabling efficient and accurate computation of the dynamical evolution of binary systems. Built upon the Framework for Developing Parallel Particle Simulation Codes (FDPS) \citep{69-Iwasawa.2016,70-Iwasawa.2020,71-Namekata.2018}, \textsc{petar} supports multi-core parallel computing and allows for a binary fraction in star clusters to reach 100\%, facilitating investigations into the evolution of binary fractions within star clusters.

Additionally, \textsc{petar} uses the updated version of the single-star and binary-star evolution codes, SSE and BSE, to model processes such as stellar wind mass loss, stellar type transitions, mass transfer, and binary mergers \citep{72-Hurley.2000,73-Hurley.2002,74-Banerjee.2020}. \textsc{petar} also incorporates the \textsc{galpy} code \citep{75-Bovy.2015} to simulate the influence of the galactic potential on cluster evolution. In our simulations, we adopted the \text{MWPotential2014} model from \textsc{galpy}, which is a simplified representation of the Milky-Way potential, comprising contributions from the bulge, disk, and dark matter halo.

\subsection{Observational samples of open clusters} \label{sec:2.2}

Based on the catalog of 7,167 open clusters published by \citetalias{47-Hunt.2023}, we select a subset of reliable clusters for numerical modeling. Firstly, we select isolated clusters that experience minimal interaction with others over their lifetimes. In this case, it is easier to locate their birthplaces in the Milky Way, thereby simplifying our analysis. To find the isolated clusters, we perform orbital traceback for all 7,167 cluster centers back to the time of their formation. Using this method, we can determine whether interactions between clusters have occurred, and also obtain the initial position and velocity of each cluster in the Milky Way. 
For the orbital traceback, all clusters are included in a single simulation in the Milky Way potential, and each is represented by a center-of-mass (CM) particle. The CM particles are initialized with coordinates ($x, y, z$) and reversed velocities ($-v_x, -v_y, -v_z$) from their present-day states in \citetalias{47-Hunt.2023}. The simulation is then integrated backward to each cluster's age to recover its initial position and velocity.
In practice, all clusters are integrated backward together up to the age of the oldest system (about $9.5$ Gyr) in the sample, and the initial conditions for younger clusters are then extracted at their corresponding ages along this integration.

To identify isolated clusters that underwent minimal inter-cluster interactions during orbital traceback, we established two sets of models differing in particle mass. In the mass-weighted set, each CM particle is assigned a total cluster mass of $M = N \cdot 0.6 M_\odot$, where $N$ is the observed number of stars in the cluster and 0.6 reflects the mean stellar mass from the canonical initial mass function (IMF) \citep{55-Kroupa.2001}. Here, inter-cluster interactions can influence CM orbits. In the massless set, each CM particle has a negligible mass ($10^{-6} M_\odot$), effectively removing inter-cluster interactions so that only the Galactic potential governs the motion. By comparing particle coordinates between the two sets, clusters with minimal positional differences are identified as isolated, while those with large discrepancies indicate significant inter-cluster interactions.  

\begin{figure}[h]
    \centering
    \includegraphics[width=1\columnwidth]{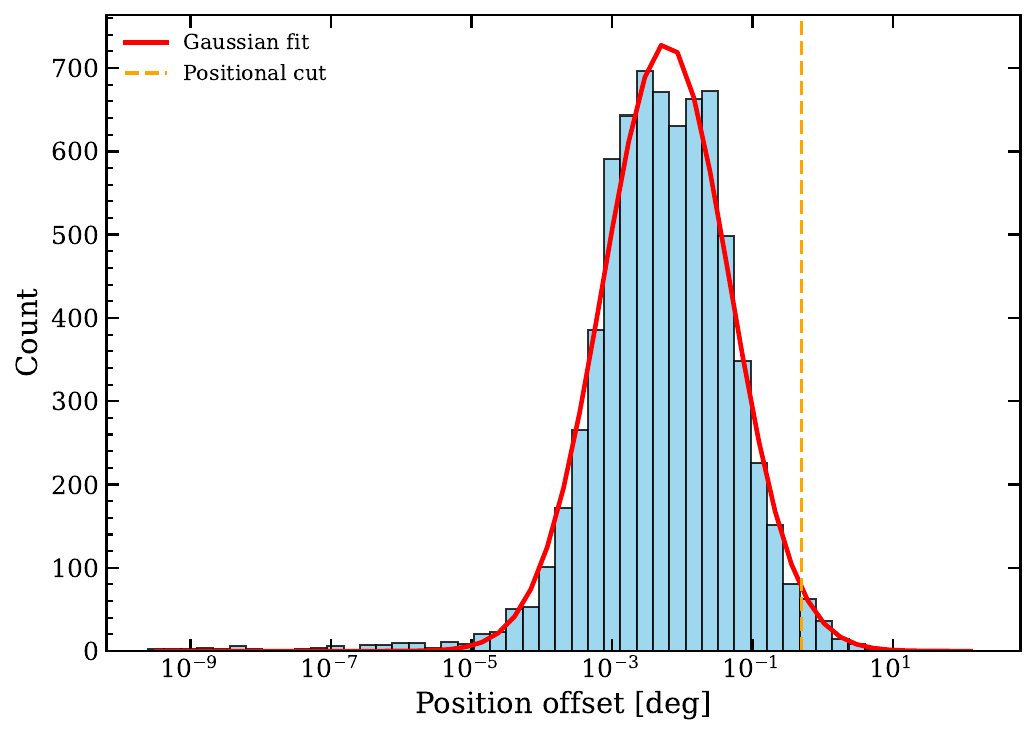}
    \includegraphics[width=1\columnwidth]{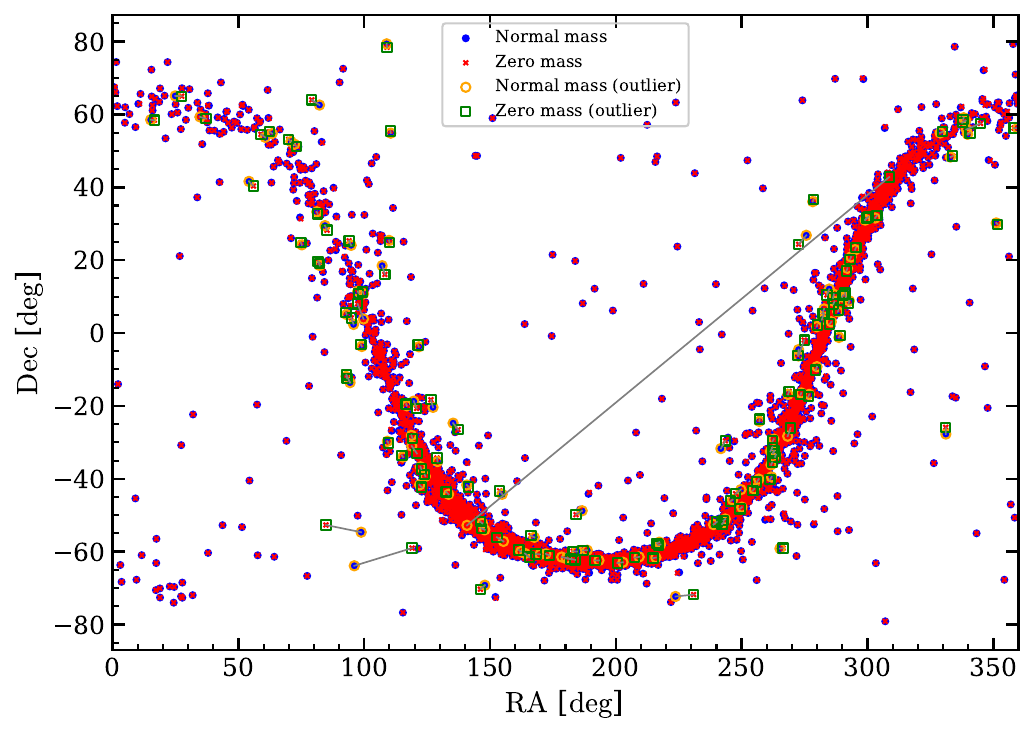}
    \caption{
    Upper panel: Histogram of positional offsets between the normal-mass and massless sets after orbital traceback for all CM particles. The offsets are binned into 50 bins and shown by the blue histogram. The red curve represents a Gaussian fit to the distribution. The vertical dashed line marks the adopted positional offset threshold of $0.5^\circ$, corresponding to approximately the 98th percentile of the empirical distribution.
    Lower panel: Spatial distribution of CM particle positions after orbital traceback. Blue dots denote the coordinates of normal-mass CM particles, while red crosses indicate those of massless particles. Orange open circles and green open squares highlight outlier sources with positional offsets exceeding the adopted threshold ($0.5^\circ$), corresponding to the normal-mass and massless sets, respectively. Light gray lines connect the paired positions of the two mass sets.
    }
    \label{fig:pos_hist_spatial}
\end{figure}

The upper panel of Figure~\ref{fig:pos_hist_spatial} displays the results of the orbital traceback coordinate offset for all CM particles between the two sets. The histogram is strongly peaked at small offsets and is well described by a Gaussian distribution. The vast majority of clusters have positional offsets below about $0.5^\circ$, while offsets larger than this value lie in the extreme right tail of the distribution, where the Gaussian fit rapidly declines. We therefore adopt a conservative cut at $0.5^\circ$, which retains the bulk of the sample while excluding extreme outliers with anomalously large positional mismatches. The adopted positional offset cut of $0.5^\circ$ corresponds to approximately the 98th percentile of the empirical offset distribution, indicating that 121 clusters ($\sim 2\%$) are excluded.

Applying this $0.5^\circ$ threshold, the position of all CM particles after undergoing traceback between two sets are shown in the lower panel of Figure~\ref{fig:pos_hist_spatial}. 
The result shows that most CM particles exhibit very similar traceback coordinates in both mass sets, while a few CM particles display significant deviations.
These deviant CM particles correspond to the extreme-offset tail identified in the upper panel of Figure~\ref{fig:pos_hist_spatial}. Therefore, they were excluded from subsequent analysis, and 7,046 systems were left. 

We then limited the cluster samples to ages below 600 Myr, as older clusters may have experienced significant mass loss from tidal evaporation and stellar evolution, complicating the determination of accurate initial conditions. To ensure the statistical reliability of N-body simulations, we required that samples contain more than 250 member stars. This criterion reduced the sample to 468 clusters.

Next, we retained only systems classified as ``o'' (open clusters) in the \citetalias{47-Hunt.2023} catalog, excluding those labeled as ``m'' (moving groups) or ``g'' (globular clusters), resulting in a working sample of 402 clusters.

Using the birth positions and velocities obtained from the traceback procedure for the remaining clusters after the above filtering steps, 
we performed forward orbital integration for each of the remaining 402 clusters and compared the simulated coordinates with their observed positions to verify whether the clusters can reach their present-day locations.
We found that the positional offsets of most samples were within $3^\circ$, and they could also be well described by Gaussian distribution. Therefore, we adopted a threshold of $3^\circ$, corresponded to approximately the 98th percentile of the positional offset distribution. Clusters exceeding this value were deemed unable to effectively reach the observed positions through simulation and were unsuitable for subsequent simulations.
This excluded 10 clusters, leaving 392 clusters.

Additionally, we excluded clusters younger than 20 Myr, since the turning point between the two stages of binary disruption typically occurs within the first $\sim$ 20 Myr. Including younger clusters would therefore sample systems that have not yet experienced this transition, making the disruption rate difficult to characterize in a uniform manner
(see Section \ref{sec3.1} for details). In total, 56 clusters were excluded at this stage. Finally, we selected 336 open cluster samples from the \citetalias{47-Hunt.2023} catalog for simulation.

\subsection{Initial condition} \label{section 2.3}
We used \textsc{mcLuster} \citep{65-Kupper.2011,Wang2019} to generate the initial conditions of star clusters. Stellar masses were optimally  sampled from the classical IMF \citep{55-Kroupa.2001} over a mass range of 0.08 to 150~$M_\odot$.
Optimal sampling fixes the number of massive stars for a given total mass \citep{Kroupa2013}, reducing stochastic variations in the evolution of low-mass clusters \citep{Wang2021}.
The initial density profile followed a spherically symmetric Plummer model \citep{83-Plummer.1911}.

To produce star cluster models with the initial total mass $M_0$ consistent with observations, we initialized $M_0 = N \cdot 0.75 M_\odot$ for each cluster, and refined it iteratively via a series of simulations using the bisection method. Each iteration minimized the difference in the number of stars within the tidal radius $\nt$ between models and observations, adopting the tidal radius $r_{\mathrm{t}}$ from \citetalias{47-Hunt.2023}. The iterations continued until the difference was less than 30, yielding the optimized mass ($M_\mathrm{opt}$).
For consistent star counts, we used \textsc{galevnb} \citep{63-pang.2016} to convert simulation snapshots into Gaia photometric data, selecting stars within the Gaia DR3 magnitude range of 4-21 in the $G$, $G_{BP}$ and $G_{RP}$ bands \citep{61-Gaia.2021,62-Gaia.2023}. To reduce computational cost, all iterations excluded primordial binaries.

To avoid the complexity of finding a proper initial $r_\mathrm{h}$ ($r_\mathrm{h,0}$) intogether with $M_0$, we adopt the $M_0-r_\mathrm{h,0}$ relation from \cite{64-Marks.2012}, given by
 $r_\mathrm{h,0}/\mathrm{pc} = 0.10^{+0.07}_{-0.04} \times (M_\mathrm{ecl}/M_\odot)^{0.13\pm0.04}$.
For our samples, the resulting $r_\mathrm{h,0}$ (\citetalias{64-Marks.2012}) values are typically below 0.25 pc.

\begin{figure}[htbp]
\begin{center}
\includegraphics[width=1\columnwidth]{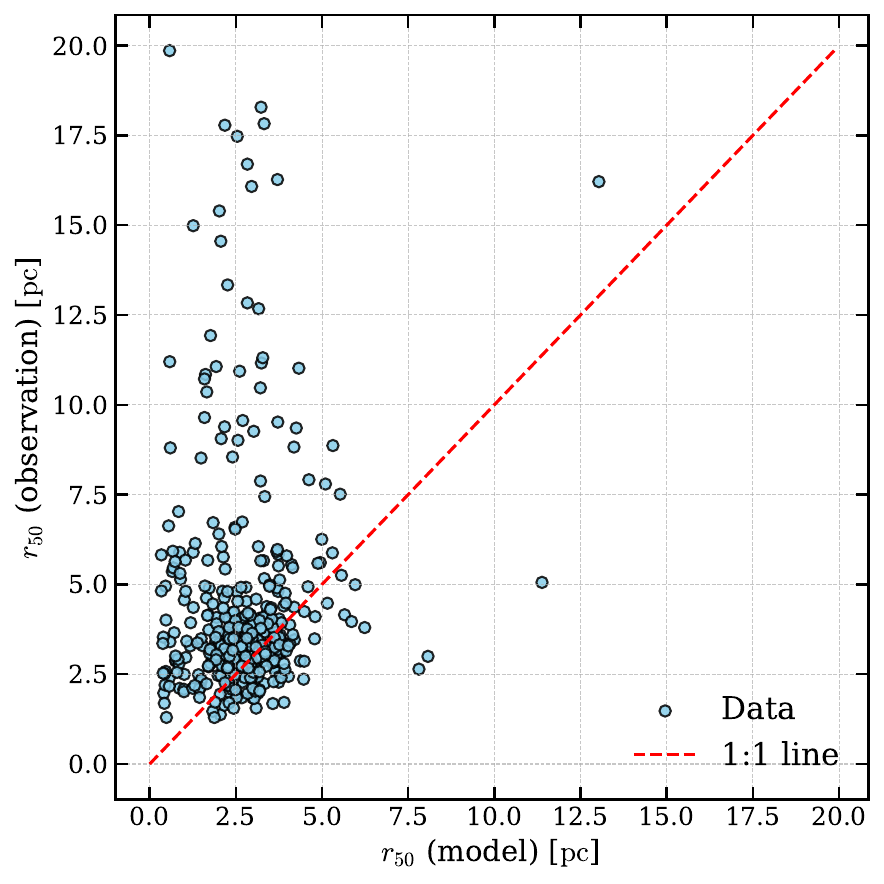}
\end{center}
\caption{Comparison of 
projected half-number radius ($r_{50}$) between simulation and the \citetalias{47-Hunt.2023} observed Catalog. The horizontal axis represents the radius containing 50\% of members within the tidal radius ($r_{50}$) of the simulated clusters, and the vertical axis represents the corresponding $r_{50}$ in the \citetalias{47-Hunt.2023} catalog. The red dashed line is a one-to-one reference line, where points closer to the line mean better agreement between simulated and observed $r_{50}$ values.}
\label{Nt_comparsion}
\end{figure}

After several iterations, our simulations well reproduce the observed $\nt$. Note that matching $\nt$ alone does not guarantee a good fit to the density profile. 
The $M_0-r_\mathrm{h,0}$ results in dense initial conditions, so the present-day $\rh$ values from simulations are often smaller than observed.
Figure \ref{Nt_comparsion} compares the observed and simulated samples in terms of the projected half-number radius, $r_{50}$, defined as the sky-plane (RA$-$Dec) radius enclosing 50\% of cluster members within the tidal radius, where members are ranked by their projected distances from the cluster center. This projected metric is adopted for consistency with the observational sample, since \citetalias{47-Hunt.2023} did not provide the three-dimensional half-mass radius $r_\mathrm{h}$.
This work aims to study binary disruption in open clusters under these conditions and to quantify the efficiency of binary disruption in the early evolution of open clusters, as well as its dependence on parameters such as $\rho_\mathrm{0}$, $P$, and $q$. Due to a small $r_\mathrm{h,0}$, the above models often have very high $\rho_\mathrm{0}$, leading to a lack of samples in low-density regimes. 
Therefore, to investigate the early binary disruption process in clusters across a broader density range and to make the conclusions more generalizable, we configured three additional $r_\mathrm{h,0}$ values: 0.5 pc, 0.8 pc, and 1.0 pc, to obtain clusters with lower $\rho_\mathrm{0}$. Due to the change in $r_\mathrm{h,0}$, the $M_0$ obtained through iterations is no longer applicable. For simplicity, we set $M_0 = N \cdot 0.75 M_\odot$, following the initial guess of the \citetalias{64-Marks.2012} models. This value is larger than the present-day mass estimate $M_0 = N \cdot 0.6 M_\odot$ (section \ref{sec:2.2}), as it approximately accounts for mass loss due to stellar evolution and tidal stripping, and should be regarded as a rough guess rather than a precise reconstruction of the initial mass.

Using the initial $M_0$ and $r_\mathrm{h,0}$ relation obtained from the iterations, we run new simulations that include primordial binaries. 
For low-mass binaries, the initial distributions of $P$, $q$, and $e$ follow \cite{29-Kroupa.1995,34-Kroupa.1995}, which assume all stars form in binaries (named as Kroupa binary model). This model contains many wide binaries that are dynamically disrupted, reproducing the observed field binary population. For high-mass binaries ($M > 5M_\odot$), we adopt the observational constrained distribution from \cite{9-Sana.2012}, which provide tighter orbits, higher $q$, and generally lower $e$.



Considering that the number of binary disruptions is particularly high within the first 10 Myr, we increased the snapshot output interval to 0.1 Myr for the first 10 Myr. After 10 Myr, as the binary disruption effect gradually weakens, the snapshot output interval was set to 1 Myr. 

To enhance the statistical robustness, 15 simulations were performed for each cluster with different random seeds to generate masses, positions and velocities of stars.
The initial parameter settings of the models with primordial binaries are summarized in Table \ref{tab:inital_param} .

\begin{table*}[htbp]
\centering
\begin{threeparttable}
\caption{Initial conditions for star cluster simulations}
\label{tab:inital_param}

\begin{tabular}{c|c|c|c|c|c|c}
\hline 
\multirow{2}{*}{Number of OCs} & \multirow{2}{*}{$M_0$} & \multirow{2}{*}{$r_\mathrm{h,0}$} & Primordial & Number of & \multirow{2}{*}{Density profile} & \multirow{2}{*}{IMF} \\
 & & & binary fraction & random seeds & & \\
\hline 

\multirow{4}{*}{336} &$M_\mathrm{opt}$ & \citetalias{64-Marks.2012} & \multirow{4}{*}{100\%} & \multirow{4}{*}{15} & \multirow{4}{*}{\citealt{83-Plummer.1911}} & \citealt{55-Kroupa.2001} \\
\cline{2-3}
 & \multirow{3}{*}{$N\cdot0.75$} & 0.5 pc & & & & with optimal \\  
\cline{3-3}
 & & 0.8 pc & & & & sampling \\
\cline{3-3}
 & & 1.0 pc & & & & \\
\hline 
\end{tabular}

\begin{tablenotes}
\footnotesize
\item $M_\mathrm{opt}$ denotes the mass obtained through the iterative method to match the observed $\mathrm{N_t}$ from \citetalias{47-Hunt.2023};
\end{tablenotes}
\end{threeparttable}
\end{table*}

\section{Results} \label{sec:3}

\subsection{Binary Disruption Rate} \label{sec3.1}
\begin{figure}[h]
    \centering
    \includegraphics[width=1.0\columnwidth]{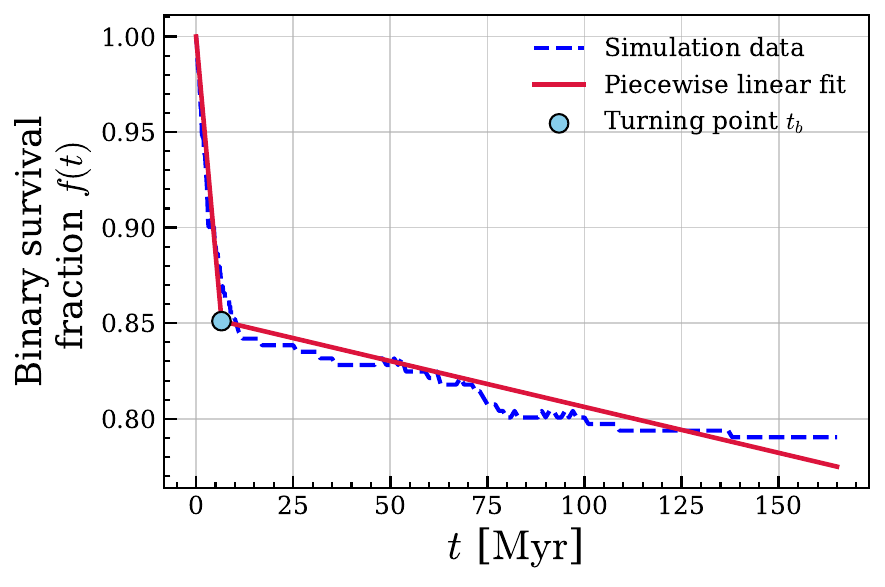}
    \caption{Example of the evolution of the binary survival fraction (Eq.~\ref{eq:ft}) for the cluster $\mathrm{NGC\_6416}$ (age 165 $\mathrm{Myr}$). The blue dashed curve shows the simulation data. The red solid line represents the piecewise linear fit described by Eq.~\ref{eq:func}, with slopes $k_1$ and $k_2$ before and after turning point $t_{\mathrm b}$. The light blue point marks the $t_{\mathrm b}$. }
    
    \label{binary_disrupt_example}
\end{figure}

By analyzing the simulations, we investigate how binaries are disrupted in each cluster model.
Binaries are identified in each snapshot using a KDTree-based nearest-neighbour search; a pair is classified as a binary if it is gravitationally bound (semi-axis $a>0$) and its apocentre distance satisfies $a(1+e)<0.1\ \mathrm{pc}$.
For each case, we track the binary count over time and define the binary survival fraction
\begin{equation}
    \label{eq:ft}
    f(t) = \frac{N_{b}(t)}{N_{b}(t_0)}
\end{equation}
as the ratio of the binary count at time $ t $ to its initial value, and its derivative representing the binary disruption rate. 
In the early stages of cluster evolution, the smaller radius and higher density of the cluster lead to frequent dynamical interactions, resulting in significant binary disruption, especially for wide binaries. As evolution progresses, the cluster's density decreases, and the average distance between stars increases, reducing the possibility of dynamical interactions. 
Consequently, $ f(t) $ exhibits a temporal turning point $t_{\mathrm{b}}$ : before this time, binaries undergo substantial disruption driven by dynamical encounters, while afterward, $ f(t) $ levels off. Statistical analysis of a large sample reveals that this turning point varies with the cluster's initial conditions, generally occurring within 20 Myr. As a result, clusters younger than this age typically remain in the early disruption phase and do not yet exhibit a well-defined turning point.
Figure \ref{binary_disrupt_example} shows an example from our simulations,
which corresponds to the cluster listed as ``$\rm NGC\_6416$" ($\rm ID:4744$) in \citetalias{47-Hunt.2023}.
The simulated cluster had an initial $r_\mathrm{h}$ of 1.0 pc and an initial mass of $297\ \mathrm{M_\odot}$.

To model the observed trend in binary disruption within star clusters, we employed a two-segment piecewise function to fit $f(t)$ . The fit function is defined as follows:

\begin{equation}
\label{eq:func}
f(t) =
\begin{cases}
  k_1 t +1 , & \text{if } t \leq t_b \\
  k_1 t_b+1+k_2(t-t_b), & \text{if } t > t_b
\end{cases}
\end{equation}

Here, $ t_{\mathrm{b}} $ represents the turning point at which $f(t)$ transitions from rapid to slow. 
The absolute value of the slope $ k_1 $ and $k_2$ represent  the binary disruption rates before and after $t_{\mathrm{b}}$. This study focuses on early binary disruption, the main phase of significant binary survival fraction change, and thus, we primarily analyze $ k_1 $ and $t_{\mathrm{b}}$, with less emphasis on $ k_2 $. 

The parameters $( k_1, k_2,t_{\mathrm b})$ were determined by minimizing the unweighted sum of squared residuals using a non-linear least-squares optimization. Initial parameter estimates were obtained from separate linear fits to the early and late evolutionary phases, and subsequently refined simultaneously via the Powell method. The goodness of fit was evaluated using the mean squared error (MSE) between the model and the simulation data.

\begin{figure}[htbp]
\begin{center}
\includegraphics[width=1.0\linewidth]{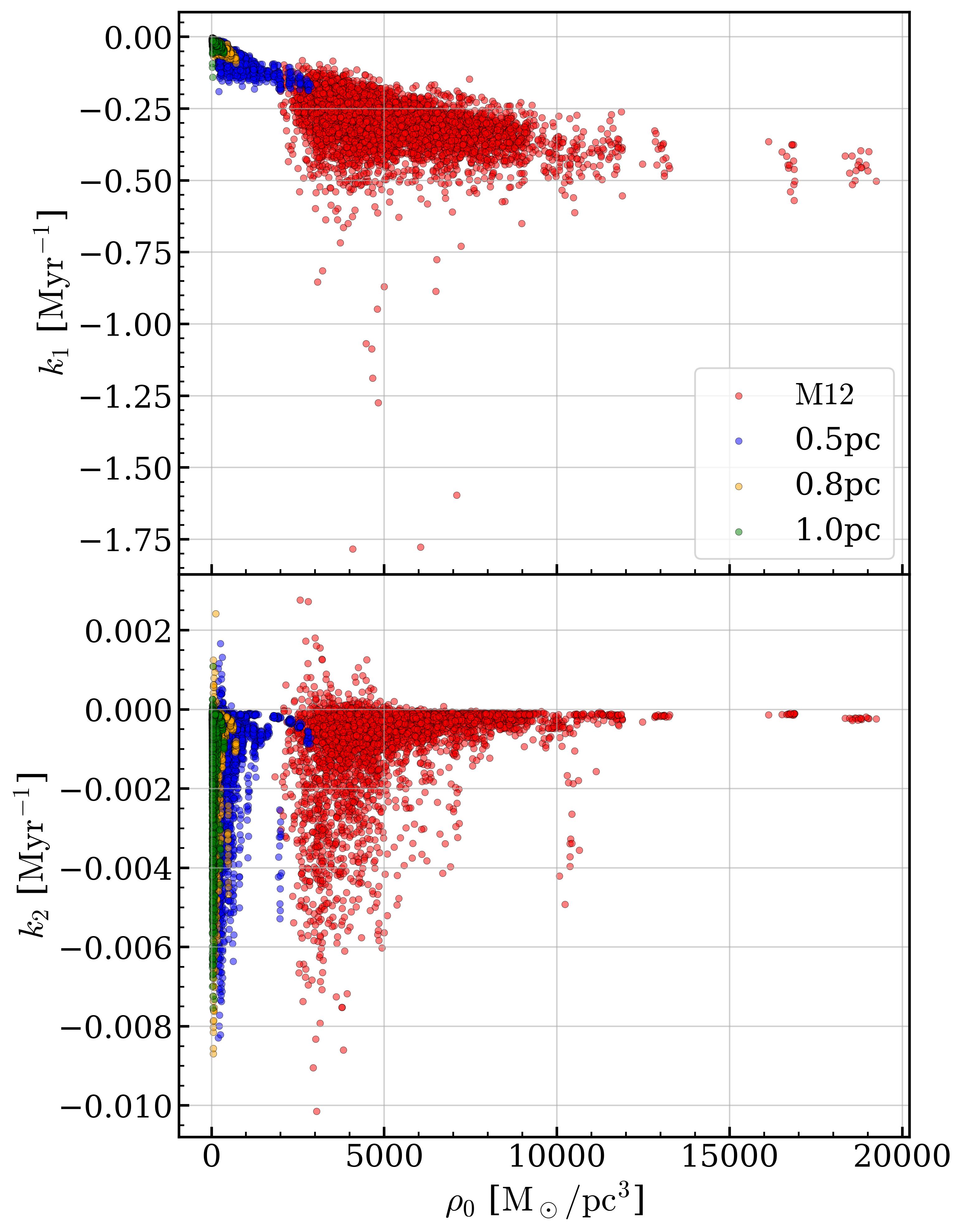}
\end{center}
\caption{Relationship between fit parameters and $\rho_\mathrm{0}$ for all samples. The upper subplot shows the distribution of $k_1$ versus $\rho_\mathrm{0}$, and the bottom subplot shows the distribution of $k_2$ versus $\rho_\mathrm{0}$. The red dots represent the sample of $ r_\mathrm{h,0} $ determined by the $ M_0 - r_\mathrm{h,0} $ relationship from \cite{60-Marks.2011}. The blue, orange, and green dots correspond to samples with $ r_\mathrm{h,0} = 0.5 \, \text{pc} $, $ 0.8 \, \text{pc} $, and $ 1.0 \, \text{pc} $, respectively.}
\label{fig:all_k1_rho}
\end{figure}

We present the distributions of $ k_1 $ and $ k_2 $ values as functions of $ \rho_\mathrm{0} $ in Figure \ref{fig:all_k1_rho}. 
The colored points denote different $r_{\mathrm{h,0}}$. The wide dispersion of $k_1$ at a given $\rho_0$ arises from statistical variations due to random seeds and cluster parameters such as $r_{\mathrm{h,0}}$. In the upper panel, the dispersion of $k_1$ increases sharply when $\rho_0 > 2500\,M_{\odot}\,\mathrm{pc}^{-3}$ and $r_{\mathrm{h,0}}$ decreases to small values from \citetalias{64-Marks.2012} (typically $\le 0.25~ \mathrm{pc}$). This suggests that frequent dynamical encounters in compact clusters cause large scatter in the binary disruption rate, while clusters with larger $r_{\mathrm{h,0}}$ (0.5, 0.8, and 1.0 $\mathrm{pc}$) experience weaker interactions and thus show smaller scatter in $k_1$. 
The uneven mass distribution of the sample also plays a role: the scarcity of massive clusters reduces the statistical representation of their stochastic effects compared to the numerous low-mass systems. A similar pattern appears in $k_2$ (lower panel of Figure \ref{fig:all_k1_rho}), where the lack of massive clusters with $r_\mathrm{h,0}=0.5-1.0\ \text{pc}$ produces a gap relative to the more abundant low-mass clusters with  \citetalias{64-Marks.2012} $r_{\mathrm{h,0}}$ at $\rho_{0} \sim 2000-2500\, M_{\odot}\ \mathrm{pc}^{-3} $. The absence of massive compact clusters further reflects this selection bias, leading to reduced dispersion of $k_2$ at the high-density end.

Overall, as $ \rho_{\mathrm{0}} $ increases, the absolute value of $ k_1 $ consistently increases, indicating that more binaries are disrupted during the early evolution. 
However, $k_2$ is generally very small compared to $k_1$, indicating that $f(t)$ remains nearly constant after $ t_{\mathrm{b}} $, which is consistent with the $f(t)$ evolution shown in Figure \ref{binary_disrupt_example}.

\begin{figure}[htbp]
\begin{center}
\includegraphics[width=1.0\columnwidth]{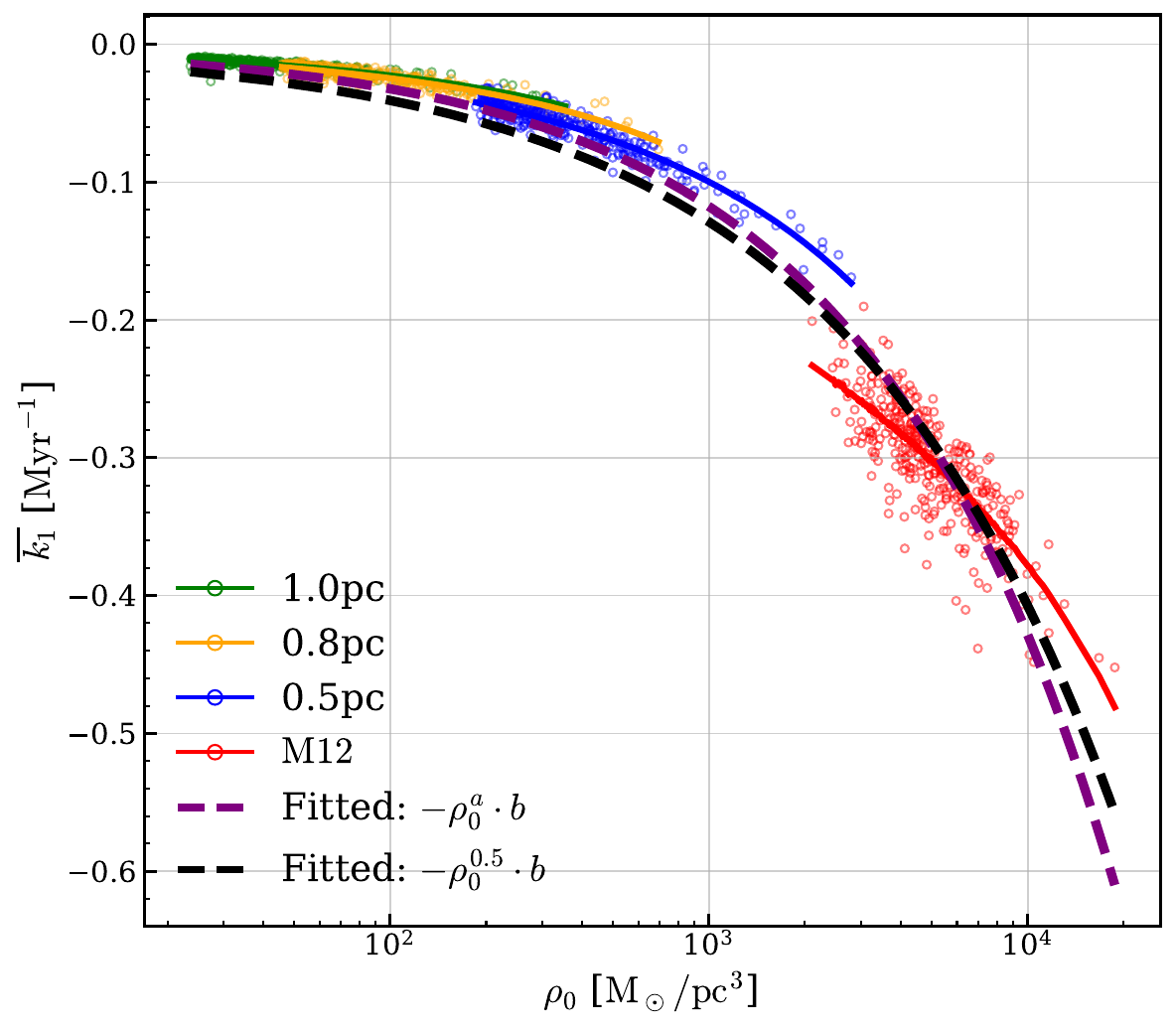}
\end{center}
\caption{Fit results for the average value of $k_1$ at different densities. The dots represent the average $k_1$ of 15 randomized models for each cluster. Different colors correspond to clusters with different  $r_{\mathrm{h},0}$ (1.0 pc, 0.8 pc, 0.5 pc, and \citetalias{64-Marks.2012}). The solid lines in the corresponding colors show the individual best-fitting results using Eq.~\ref{eq:coll_rate3}. The purple dash line shows the result of fitting all $ \overline{k_1} $ using Eq.~\ref{eq:all_k1_rho_fitting}, and the black dash line shows the result of the re-fit of $ \overline{k_1} $ after fixing the parameter $ a $ in Eq.~\ref{eq:all_k1_rho_fitting} to 0.5. }

\label{fig:mean_k1_rho_fitting}
\end{figure}

To better characterize the early binary evolution trajectories of star clusters, we calculated the averaged $k_1$ from the 15 randomized models for each cluster ($\koneave$) to quantify its mean evolutionary trajectory, as shown in the colorful dots of the Figure ~\ref{fig:mean_k1_rho_fitting}.

To interpret this result, we assume that binary disruption is driven by encounters and is characterized by the collision rate given by equation 7.194 in \citet{84-Binney.1987}:

\begin{equation}
    \frac{1}{t_\mathrm{coll}} = 4\sqrt{\pi}n\sigma \left (r^2_{\mathrm{coll}}+\frac{G m}{\sigma^2}  r_\mathrm{coll} \right)
    \label{eq:coll_rate}
\end{equation}
where $t_\mathrm{coll}$ is the collision timescale, $n$ is the stellar number density, $m$ is the stellar mass, $G$ is gravitational constant, $r_\mathrm{coll}$ is the collision radius (here approximated by the binary separation), and $\sigma$ is the cluster velocity dispersion.
Under virial equilibrium, $\sigma \approx \sqrt{GmN / r_\mathrm{h}}$ within $\rh$ and $\rh = [4 N / (3\pi n)]^{1/3}$ , yielding

\begin{equation}
    \frac{1}{t_\mathrm{coll}} =  \beta  \left( n^{\frac{3}{2}} r_\mathrm{coll}^2 r_\mathrm{h}  + \frac{n^{\frac{1}{2}}r_\mathrm{coll}}{r_\mathrm{h} } \right)
    \label{eq:coll_rate3}
\end{equation}

where $\beta$ is a constant. 
The first term in the bracket corresponds to the close encounter rate of binaries, while the latter term corresponds to the gravitational focusing effect of binaries.

The collision rate depends not only on $n$ but also on $r_\mathrm{h}$ and $r_{\mathrm{coll}}$, implying $a$ is not universal for all binaries. 
According to Eq.~\ref{eq:coll_rate3}, we therefore fit $\overline{k_1}$ versus $\rho_0$ for clusters with different $r_\mathrm{h,0}$, adopting $\beta$ (in $\rm pc ^{3/2}\ Myr^{-1}$) and $r_{\mathrm{coll}}$ (in $\rm pc$) as free parameters. The results are shown in Figure~\ref{fig:mean_k1_rho_fitting}. For clusters with $r_{\mathrm{h,0}} = 1.0$ pc, $0.8$ pc, $0.5$ pc, and \citetalias{64-Marks.2012}, the fitted $\beta$ values are $-9.5535 \pm 4.4330$, $-7.8787\pm 2.6659$, $-16.8258 \pm 7.2448$, and $-13.4357 \pm 4.8284$, respectively, while the corresponding $r_{\mathrm{coll}}$ values are $2.31 \times 10^{-4} \pm 1.05 \times 10^{-4}$, $2.46 \times 10^{-4} \pm 8.10\times10^{-5}$, $9.16\times10^{-5}\pm3.88\times10^{-5}$, and $6.86\times10^{-5}\pm2.42\times10^{-5}$. 

While this fit works well, it is overly complex and results in different parameters for different $r_{\mathrm{h,0}}$. Given the large scatter in $k_1$ (Figure~\ref{fig:all_k1_rho}), we instead adopt a simpler, more interpretable relation: $ \overline{k_1}$ exhibits a power-law relationship with $ \rho_{\mathrm{0}} $. To quantitatively characterize this relationship, we performed a fit using 
\begin{equation}
    \overline{k_1}(\rho_0) = -\rho_\mathrm{0}^a \cdot b 
    \label{eq:all_k1_rho_fitting}
\end{equation}
The fit results of our data are shown in the  Figure \ref{fig:mean_k1_rho_fitting} (purple dash line), with the fit parameters being $a = 0.5614 \pm 0.0043$ and $b = 2.434 \times 10^{-3} \pm 8.9 \times 10^{-5}$.

The fit parameter $a\approx0.5614$ 
suggests that binary disruption is mainly governed by the gravitational focusing term ($\propto n^{1/2}$) in Eq.~\ref{eq:coll_rate3}, rather than the close-encounter term ($\propto n^{3/2}$). 
Deviations from the fit curve in Figure~\ref{fig:mean_k1_rho_fitting} further reflect the influence of $r_\mathrm{h}$ and  $r_{\mathrm{coll}}$, 
which introduce secondary corrections to the primary dependence on $\rho_0$.
For the overall binary population, the presence of close binaries results in a smaller average $r_\mathrm{coll}$, weakening the influence of the first term, while the gravitational focusing effect of such binaries becomes more pronounced, leading to a direct correlation between $\overline{k_1}$ and $\sqrt{\rho_0}$. 


This density dependence is also physically consistent with the disruption timescale $t_{\mathrm{dis}}$ of wide binaries.
During the early evolution of open clusters, wide binaries are preferentially disrupted, with $ t_{\mathrm{dis}} \propto \tch$ 
\citep{60-Marks.2011},
where $\tch$ at $\rh$ is given by \cite{59-Spitzer.1987}: 
\begin{equation}
    \tch = \frac{\rh^{{3/2}}}{(GM)^{1/2}}\propto \frac{1}{(G\rho)^{1/2}}
    \label{crossing_time}
\end{equation}
where $M$ is the total mass and $\rho$ is stellar density at $\rh$. Thus,
\begin{equation}
    t_\mathrm{dis} \propto \frac{1}{\sqrt{\rho_\mathrm{0}}}
    \label{eq:tids_rho}
\end{equation}
where $\rho$ is approximated by its initial value.
Since $\koneave$ represents the binary disruption rate, 
\begin{equation}
    \koneave(\rho_0) \propto - \frac{1}{t_\mathrm{dis}} \propto -\sqrt{\rho_\mathrm{0}}
    \label{eq:k1_tcr}
\end{equation}
Therefore, after fixing the parameter $ a = 0.5$, we obtained new fit results shown as the black dash line in Figure \ref{fig:mean_k1_rho_fitting}, 
with $b = 0.0040736 \pm 0.00001613 $. 
Although minor deviations exist, the new curve still reproduces the trend of $ k_1 $ with respect to $ \rho_\mathrm{0} $ reasonably well. 

\begin{figure}[htbp]
\begin{center}
\includegraphics[width=1.0\columnwidth]{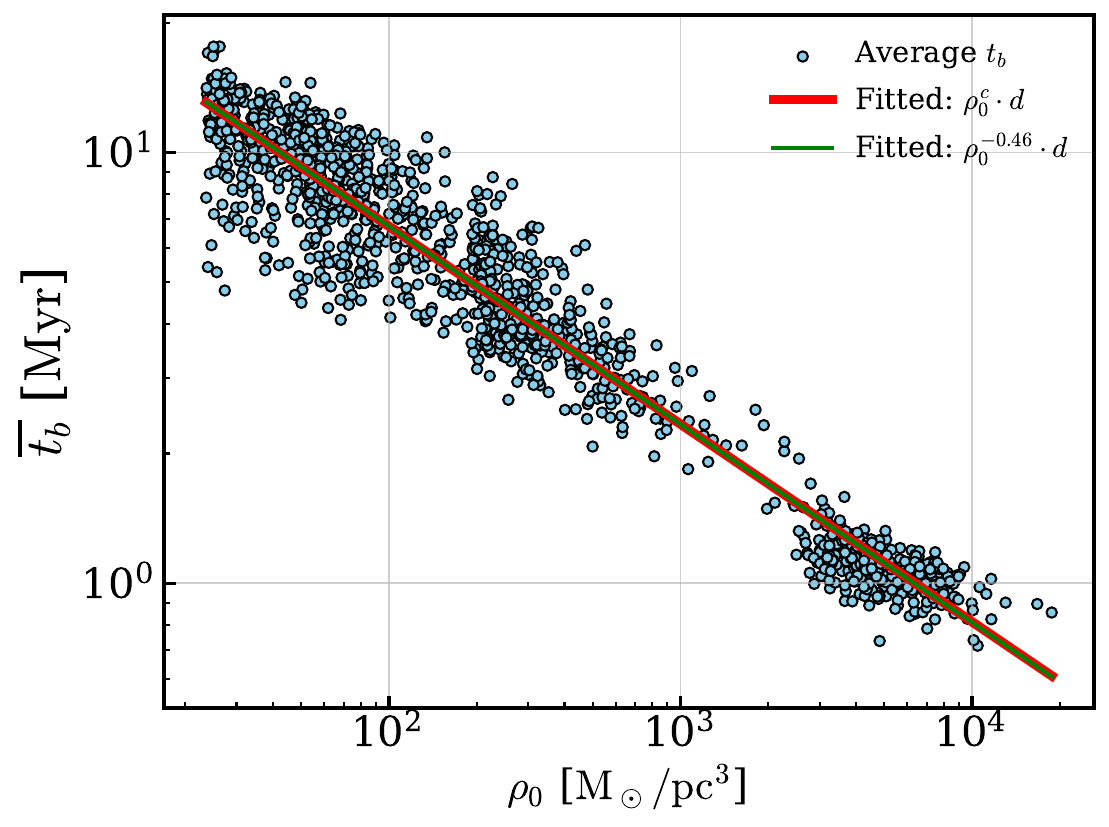}
\end{center}
\caption{Distribution and fit result for the average value of $t_{\mathrm{b}}$ with respect to $\rho_\mathrm{0}$.The blue dots represent the average $t_\mathrm{b}$ of 15 randomized models for each cluster, the red line shows the result of the fitted $ \overline{t_{\mathrm{b}}} $ using Eq.~\ref{eq:tb_rho}, and the green line represents the fit result after fixing the parameter $ c $ in Eq.~\ref{eq:tb_rho} to $-0.46$. }
\label{fig:tb_rho_fitting}
\end{figure}

For the binary disruption mean transition time over different random seeds $\overline{t_{\mathrm{b}}}$ 
in Eq.~\ref{eq:func}, we find a power-law dependence on $\rho_\mathrm{0}$. Figure \ref{fig:tb_rho_fitting} shows $t_\mathrm{b}$ versus $\rho_\mathrm{0}$ and the fit (red line)
\begin{equation}
    \overline{t_\mathrm{b}}(\rho_0) =   \rho_\mathrm{0}^{c} \cdot d,
    \label{eq:tb_rho}
\end{equation}
with $c = -0.4591 \pm 0.003163$ and $d = 55.812\pm1.0526$. 
We fixed the parameter $ c $ at $-0.46$, achieving a good fit, as shown by the green line in Figure \ref{fig:tb_rho_fitting}, with the corresponding fitting parameter $ d $ being $ 56.090 \pm 0.3322 $.

Based on this relation, we estimated $t_\mathrm{b}$ for all clusters using stellar densities inferred from the observed $r_\mathrm{50}$ reported in \citetalias{47-Hunt.2023}. Specifically, we converted the observed  $r_\mathrm{50}$ to an effective half-mass radius $r_\mathrm{h}$ and computed the corresponding stellar density, which was then used to evaluate $t_\mathrm{b}$.
The results show that approximately two-thirds of clusters satisfy $t_\mathrm{b}$ less than observed ages ($t_\mathrm{obs}$). The remaining clusters that do not meet this criterion are primarily those with relatively large observed radii, which correspond to lower stellar densities and therefore yield larger values of $t_\mathrm{b}$.
We note, however, that the adopted stellar densities are based on present-day parameters rather than the initial density $\rho_0$. Since clusters are expected to expand over time due to gas expulsion, stellar evolution mass loss, and dynamical heating, the initial half-mass radius $r_\mathrm{h,0}$ were likely smaller and the corresponding  $\rho_0$ higher than the values inferred from the current  $r_\mathrm{50}$. Therefore, our estimates of  $t_\mathrm{b}$ are likely conservative, and the actual fraction of clusters satisfying  $t_\mathrm{b} < t_{\mathrm{obs}}$  may be even larger.

\begin{figure}[htbp]
\begin{center}
\includegraphics[width=1.0\columnwidth]{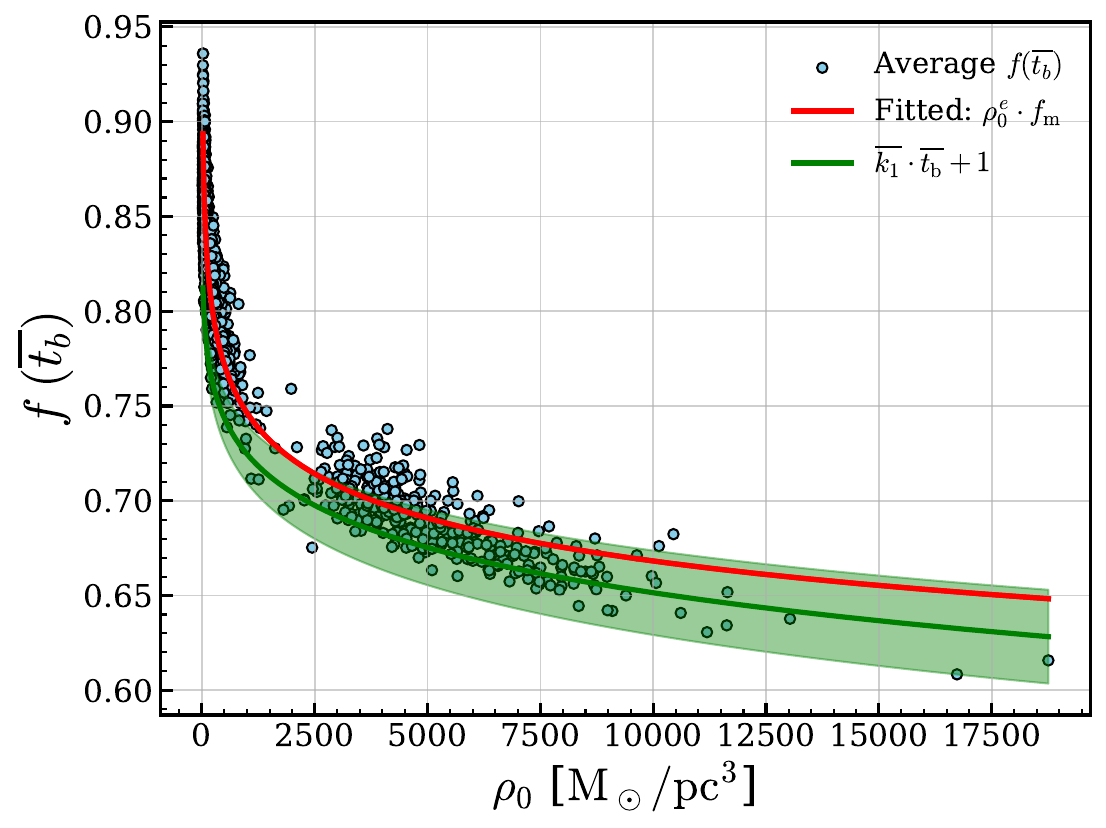}
\end{center}
\caption{Distribution of the average value of $ f(t_\mathrm{b}) $ with respect to $ \rho_\mathrm{0} $ and the corresponding fit results. The blue dots represent the average value of $ f(t_\mathrm{b}) $ at the moment $ t_{\mathrm{b}} $ for 15 randomized models of each cluster. The red line shows the result of $ f(\overline{t_\mathrm{b}}) $ from Eq.~\ref{eq:ftb_rho}. The green line represents $ f(\overline{t_\mathrm{b}}) $ at the moment $ t_{\mathrm{b}} $, expressed using $ \overline{k_1} $ and $ \overline{t_\mathrm{b}} $ according to Eq.~\ref{eq:func}, with the green band indicating the corresponding uncertainty.}
\label{fig:fb_rho_fitting}
\end{figure}

According to Eq.~\ref{eq:func}, the binary survival fraction $f(\overline{t_\mathrm{b}})$ at time $\overline{t_\mathrm{b}}$ can be expressed using both $\overline{k_1}(\rho_0)$ and $\overline{t_\mathrm{b}}(\rho_0)$.
Figure \ref{fig:fb_rho_fitting} shows the relationship between $f(\overline{t_\mathrm{b}})$ and $\rho_\mathrm{0}$, along with the estimated values derived from $\overline{k_1}(\rho_0)$ and $\overline{t_\mathrm{b}}(\rho_0)$ with uncertainties (green region). Most data points fall within this uncertainty range, though some lie above it, likely due to anomalies in the $\koneave(\rho_0)$ fit.

We further fit $f(\overline{t_\mathrm{b}})$ as a function of $\rho_\mathrm{0}$ using Eq.~\ref{eq:ftb_rho} (red line), 
\begin{equation}
    f(\overline{t_\mathrm{b}},\rho_0) =   \rho_\mathrm{0}^{e} \cdot f_{\rm m},
    \label{eq:ftb_rho}
\end{equation}
which matches the green region. The best-fit parameters are 
$e = -0.04805 \pm 0.00035$ and $f_{\rm m} = 1.0402 \pm 0.0022$.
The results indicate a weak power-law relationship between $f(\overline{t_\mathrm{b}})$ and $\rho_\mathrm{0}$, with $f(\overline{t_\mathrm{b}})$ exhibiting a soft lower limit as $\rho_\mathrm{0}$ increases. In other words, for a given open cluster, there is an upper limit to $f(t)$. 

\subsection{Relationship Between the Binary Disruption Rate and Binary Parameters} \label{sec3.2}

\subsubsection{Dependence on Mass Ratio} \label{sec3.2.1}

\begin{figure*}[!t]
    \centering
    \includegraphics[width=1\columnwidth]{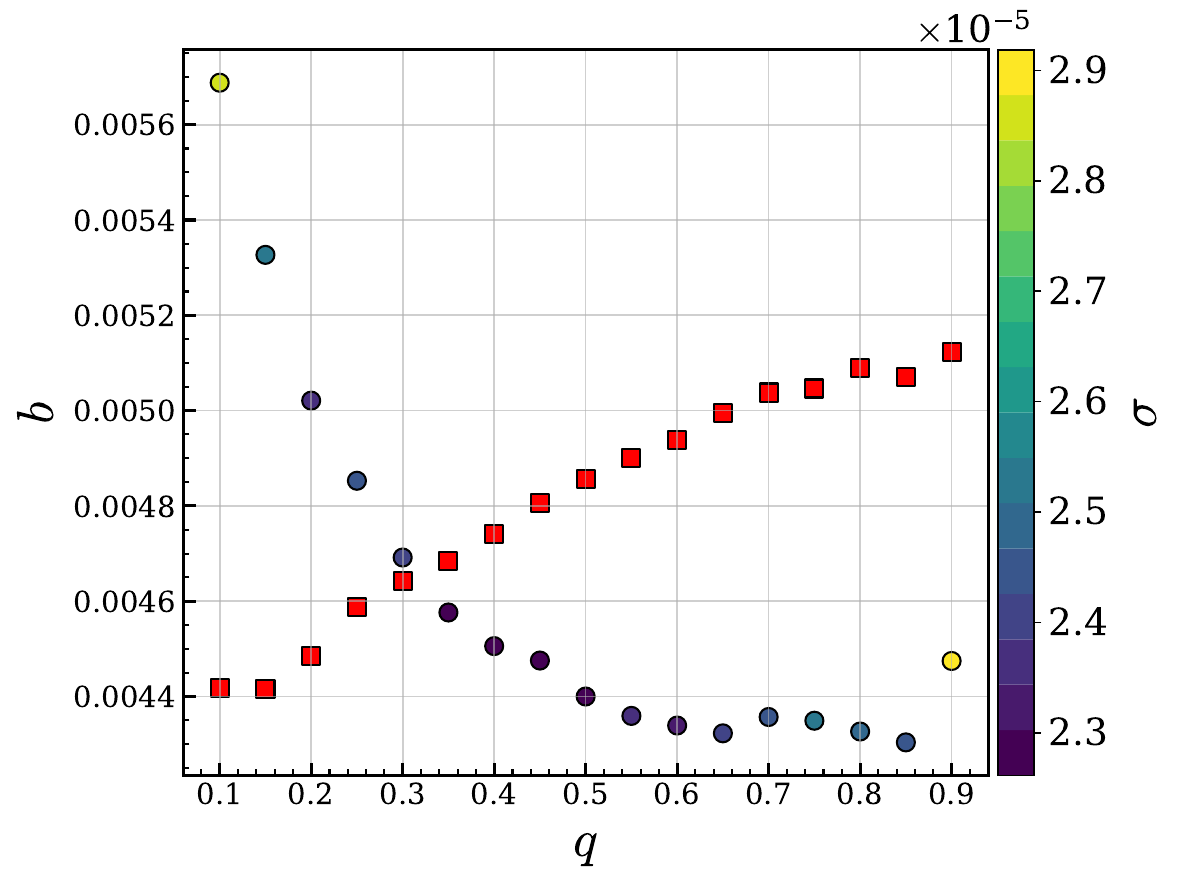}
    \includegraphics[width=1\columnwidth]{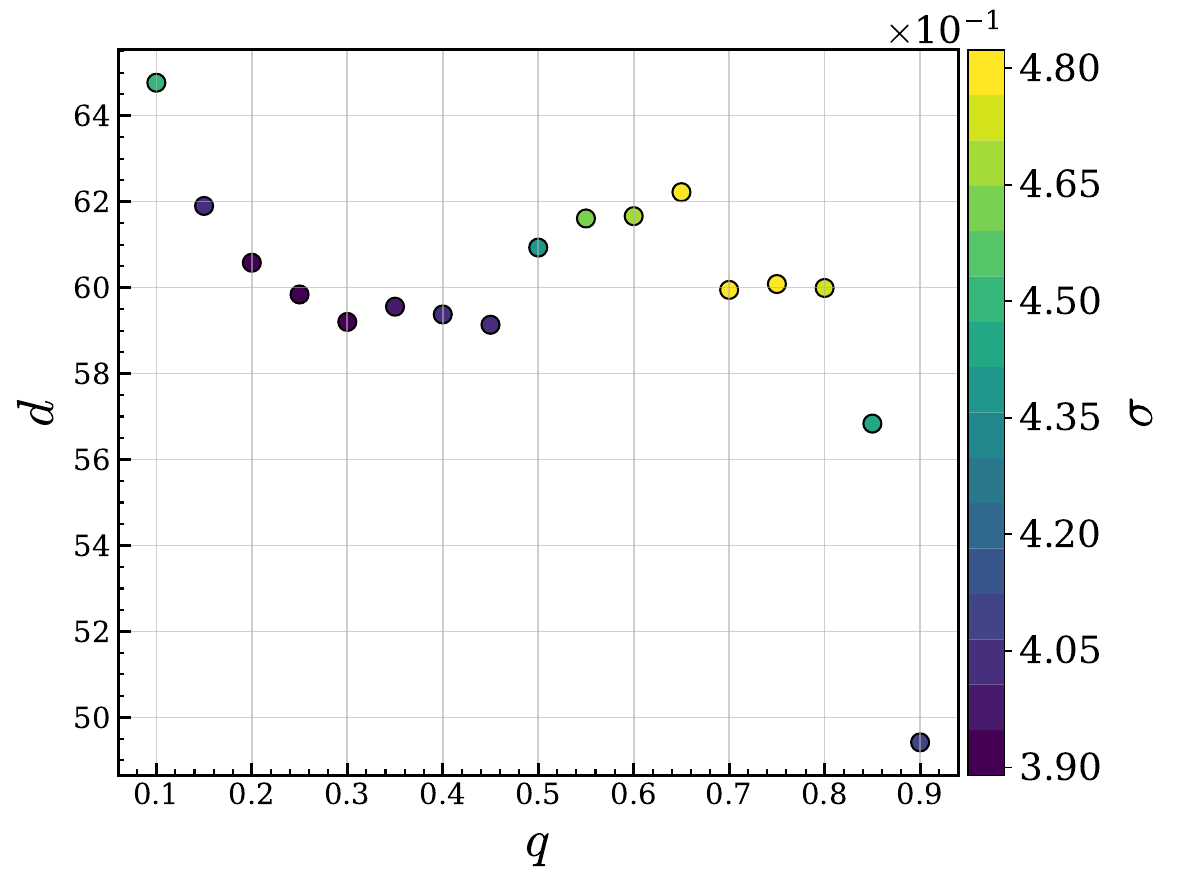}
    \caption{
    The relationship between the fit parameters $ \overline{k_1} $, $ \overline{t_\mathrm{b}} $, and $ q $. The left subplot shows the relationship between the fitted parameter $ b $ and $ q $ after fixing the parameter $ a $ (i.e., $0.5$) in Eq.~\ref{eq:all_k1_rho_fitting}, and red squares represent the fit result that excluded short period binaries ($\log_{10}(P/\text{day}) < 5$).
    The right subplot shows the relationship between the fitted parameter $ d $ and $ q $ after fixing the parameter $ c $ (i.e., $-0.46$) in Eq.~\ref{eq:tb_rho}. The horizontal axis of both subplots represents the median value of each $ q $ interval, while the vertical axes represent the parameters $ b $ and $ d $, respectively. The color bars indicate the fit uncertainties for each parameter.}
    \label{fig:kb_td_q}
\end{figure*}

We investigated the impact on $ q $ of $f(t)$ . To obtain reliable statistics, we employed a sliding window method to divide $ q $ into multiple intervals: each interval had a width of $ \Delta q = 0.2 $, starting from $ q = 0.0 $, with a sliding step of 0.05, resulting in 17 intervals. At each time step, we tracked the binary population within each $ q $ interval. Across all intervals, it exhibits a two-stage evolution, consistent with the global trend shown in Figure~\ref{binary_disrupt_example}. Accordingly, following Section \ref{sec3.1}, we fit each interval's evolution using Eq.~\ref{eq:func} and obtain  $ \koneq $ and $\tbq$ values for each $ q $ interval.
We then analyzed their relationship with $\rho_{\mathrm{0}}$ and find that they still follow a power-law distribution, as described by Eq.~\ref{eq:all_k1_rho_fitting} and \ref{eq:tb_rho}. By fitting Eq.~\ref{eq:all_k1_rho_fitting} to the mean $\koneq$ across 15 random-seed models ($\koneave(q,\rho_0)$), the parameters $a(q)$ and $b(q)$ exhibited some coupling, indicating overfitting.
To mitigate this, we fixed the value of $ a(q) $ at 0.5, as mentioned in Section \ref{sec3.1}.
Then we refitted $b(q)$ to simplify the relationship between $\koneave(q,\rho_0)$ and $ q $, making the variation trend clear while maintaining a good fit accuracy. Using the same approach for $ \overline{t_\mathrm{b}}(q,\rho_0) $, we fixed $ c(q) $ in Eq.~\ref{eq:tb_rho} to $-0.46$ and fitted $ d(q) $.
The fit parameters are shown in Figure \ref{fig:kb_td_q}. 

As shown in Figure \ref{fig:kb_td_q}, the parameter $ b(q) $ displays a clear inverse correlation with $ q $, indicating that binaries with smaller $ q $ are more easily disrupted at the same $ \rho_{\mathrm{0}} $. 
However, this relation includes binaries of all orbital periods. 
When binaries with $\log_{10}(P/\mathrm{day}) < 5$ are excluded and the data are refitted, the correlation reverses: $b(q)$ increases with $q$, as shown by the red squares in Figure \ref{fig:kb_td_q}. This indicates that the anti-correlation of $b-q$ is dominated by the short-period binaries, whereas high-$q$ wide binaries are more easily disrupted.

To understand this phenomenon, 
we track the evolution of initial short-period binaries. A binary is classified as dynamically disrupted if it becomes unbound in subsequent snapshots, while it is identified as a merger if the two stellar components coalesce into a single object. Among the short-period binaries that disappear over time, about 60\% results from dynamical disruption, while roughly 40\% arises from mergers. 
The dynamical disruption as a function of $q$ follows a trend similar to that of wide binaries, while mergers are more common in low-$q$ and $q=1$ short-period binaries, with the latter arising from the eigenevolution of the primordial binary model \citep{34-Kroupa.1995}. 

In Figure \ref{fig:kb_td_q}, $ d $ remains nearly constant at low $ q $ but decreases rapidly as $ q $ approaches 1. 
Unlike $b(q)$, we do not observe any dependence of this trend on orbital period, suggesting that the variation of $\tbave$ with $q$ is largely independent of binary separation. 
This indicates that wide binaries with more similar masses disrupt more quickly under identical $ \rho_\mathrm{0} $. These results allow $\koneave$ and $ \tbave $ to be evaluated for any given $ \rho_\mathrm{0} $ and $ q $.


\subsubsection{Dependence on periods}
\label{sec:period}

We applied the same methodology as for $\koneave (q,\rho_{\mathrm{0}})$ to analyze the relationship among $ \koneave $, $ \rho_{\mathrm{0}} $, and $P$, denoted as $\koneave(P,\rho_0)$. During early cluster evolution, short-period binaries tend to survive, while long-period binaries with $\log_{10}(P/\text{day}) > 5$ undergo significant disruption \citep{44-Heggie.1975,45-Hills.1975}. A small fraction of very short-period binaries may also merge. The binary periods in the simulations span approximately $ -1 < \log_{10}(P/\text{day}) < 9 $. Focusing on the long-period binary disruption, we divided $ \log_{10}(P) $ into 21 sliding intervals of width $ \Delta \log_{10}(P/\text{day}) = 2 $, starting from $ \log_{10}(P/\text{day}) =5$ to 9 in sliding steps of 0.1. 
Within each interval, $\koneave(P,\rho_0)$ and $\overline{t_\mathrm{b}}(P,\rho_0)$ both follow power-law relationships  with $\rho_0$ , which can be fitted using Eq.~\ref{eq:all_k1_rho_fitting} and \ref{eq:tb_rho}.

\begin{figure*}[!t]
    \centering
        \includegraphics[width=1\columnwidth]{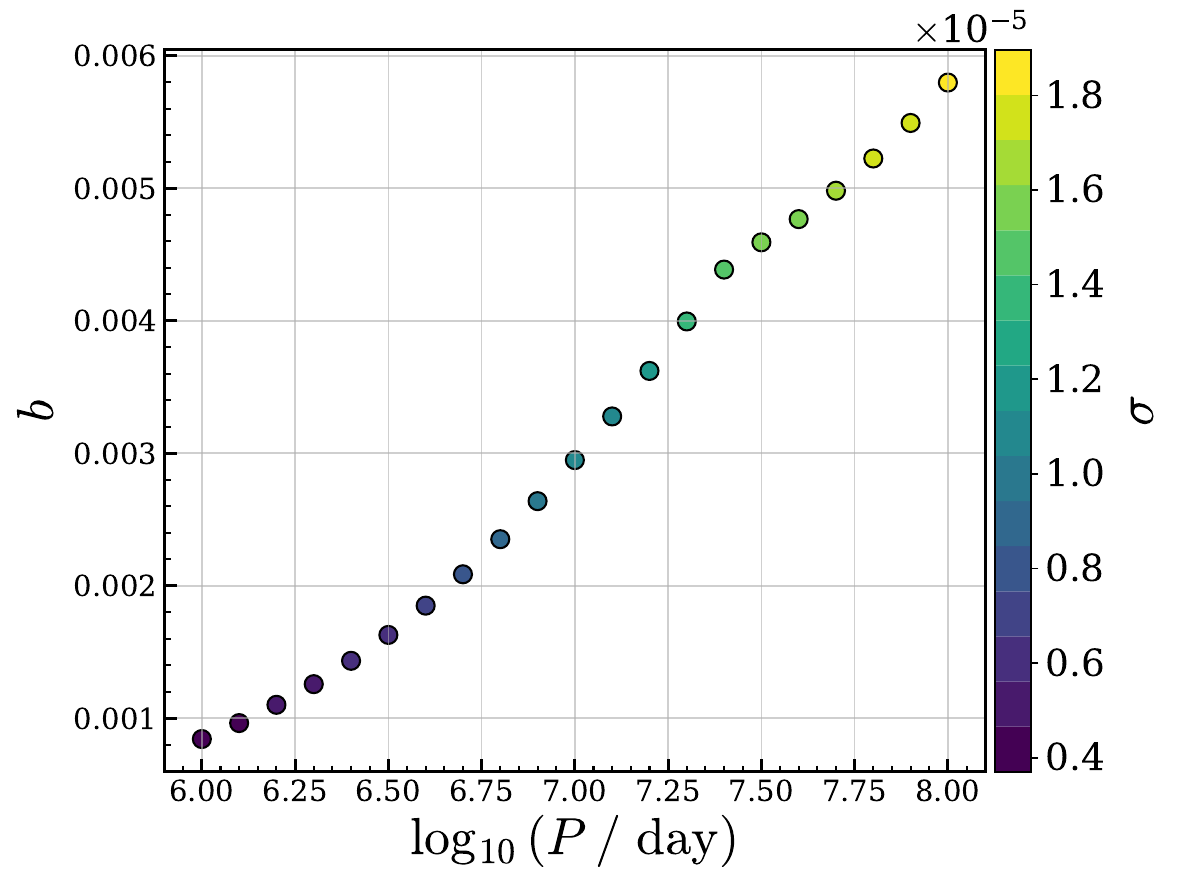}
    \hfill
        \includegraphics[width=1\columnwidth]{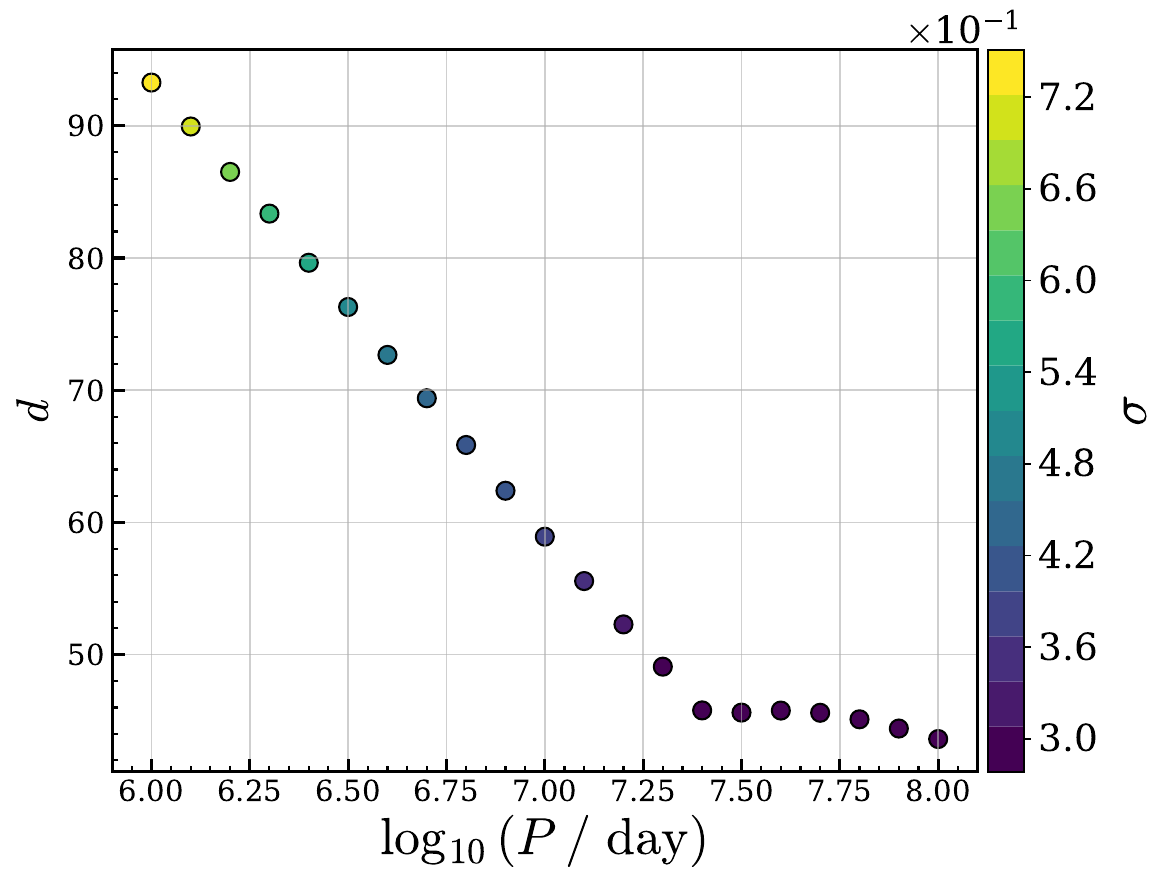}

    \caption{The relationship between the fit parameters $ \overline{k_1} $, $ \overline{t_\mathrm{b}} $, and $ \log_{10}(P) $. The left subplot shows the relationship between the fitted parameter $ b $ and $ \log_{10}(P) $ after fixing the parameter $ a $ as the mean value of $a(P)$ (i.e., $0.6348$) in Eq.~\ref{eq:all_k1_rho_fitting}. The right subplot shows the relationship between the fitted parameter $ d $ and $ \log_{10}(P) $ after fixing the parameter $ c $ (i.e., $-0.46$) in Eq.~\ref{eq:tb_rho}. The horizontal axis of both subplots represents the median value of each $ \log_{10}(P) $ interval, while the vertical axes represent the parameters $ b $ and $ d $, respectively. The color bars indicate the fit uncertainties for each parameter.}
    \label{fig:kb_td_p}
\end{figure*}

Unlike the case of $\koneave(q,\rho_0)$, the relation between $\koneave(P,\rho_0)$ and $\rho_0$ cannot be well reproduced by fixing $a=0.5$ in Eq.~\ref{eq:all_k1_rho_fitting}. Therefore, we allowed both $a$ and $b$ to vary and obtained their best-fit values, $a(P)$ and $b(P)$, for each $P$ bin. The fitted $a(P)$ values range from 0.61082 to 0.64994. 
However, due to possible overfitting in some period ranges, the variation of $a(P)$ does not fully adhere to this trend. To mitigate this effect, we adopted the mean value $\langle a(P) \rangle = 0.6348$, as a fixed exponent, which yields a significantly improved overall fit.
The deviation of $a(P)$ from 0.5 for wide binaries can be explained by Eq.~\ref{eq:coll_rate3}. Because binaries have larger $r_\mathrm{coll}$, the first term dominates, shifting the dependence of $1/t_\mathrm{coll}$ on $n$  from $1/2$ toward $3/2$. 
As illustrated in Figure \ref{fig:kb_td_p}, the fit results indicate that $ b(P) $ increases significantly  with $ \log_{10}(P) $, consistent with the faster disruption of wide binaries.

We fit $\overline{t_\mathrm{b}}(P)$ following Section \ref{sec3.2.1}, obtaining a consistent $c(P)=-0.46$ in Eq.~\ref{eq:tb_rho}. As shown in Figure \ref{fig:kb_td_p}, $d(P)$ displays a strong anticorrelation at small $\log_{10}(P)$, indicating that the rapid disruption phase of binaries with longer periods ends earlier. Beyond $\log_{10}(P)\sim7.5$, $d(P)$ flattens, indicating a lower limit of $\tb$ for wide binaries. 



\subsubsection{Dependence on orbital eccentricity}
\label{sec:ecc}

\begin{figure}[htbp]
\begin{center}
\includegraphics[width=1\columnwidth]{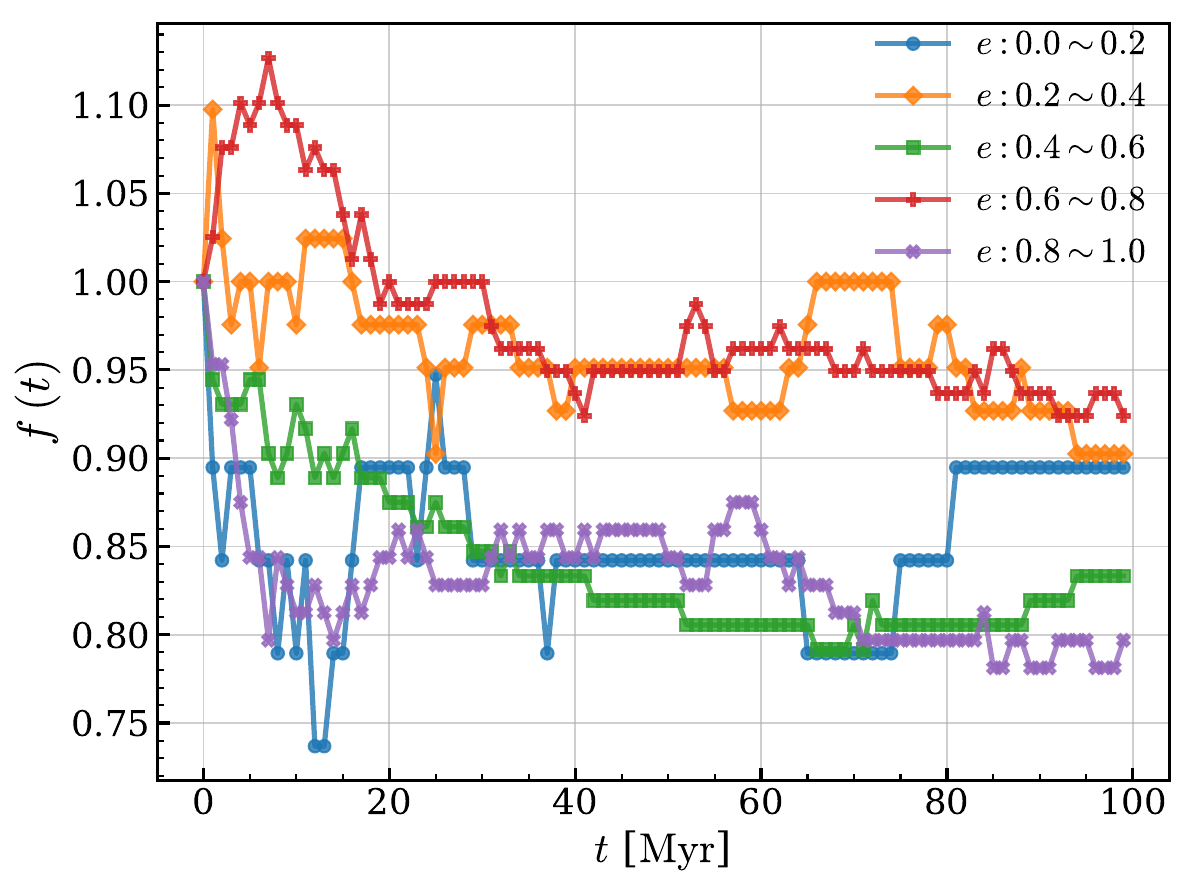}
\end{center}

\caption{Example of the time evolution of the binary survival fraction in different eccentricity ($e$) ranges for a cluster from $\rm Alessi\_24$ in \citetalias{47-Hunt.2023}, with an age of $99\ \rm Myr$. The blue, orange, green, red, and purple curves correspond to binaries with $e=0.0-0.2$, $0.2-0.4$, $0.4-0.6$, $0.6-0.8$, and $0.8-1.0$, respectively.}
\label{fig:ecc_example}
\end{figure}

In addition to $q$ and $P$, $f(t)$ also depends on eccentricity $e$, though establishing a clear relation is difficult becasue $e$ varies rapidly through dynamical interactions.
We applied the same sliding-bin method for $e$ as for the $q$ and $P$ analyses. However, frequent transitions of binaries between $e$ intervals complicated the evolution trend, causing $f(t)$ to occasionally exceed 1 and fluctuate strongly, making it challenging to derive a simple fit relation. 
Figure~\ref{fig:ecc_example} shows $f(t)$ for different eccentricity $e$ intervals (from 0.0 to 1.0, in steps of 0.2) in the cluster model $\rm Alessi\_24$, ID=82 with $r_\mathrm{h,0}=1.0\ \rm pc$ and mass $=278\ \rm M_\odot$. 
$f(t)$ initially varies significantly as binaries migrate from other $e$ intervals, then drops rapidly and fluctuates over time. As a result, Eq.~\ref{eq:func} cannot be applied to fit the data as it can for $f(t)$ in $q$ and $P$ bins.

It should be noted that this issue does not stem from biases in the statistical process itself but rather from the inherent limitations of $e$ as a highly sensitive evolutionary parameter. In contrast, while $q$ and $\log_{10}(P)$ also evolve over time, their variations are significantly less pronounced than those of $e$, allowing for better interpretability. Consequently, this study focuses solely on analyzing the dependence of $f(t)$  on $q$ and $\log_{10}(P)$, leaving the exploration of $e$-related effects for future research. 

As an alternative approach to study eccentricity evolution, we no longer bin binaries according to their instantaneous eccentricities at each time step. Instead, binaries are grouped based on their initial eccentricity $e_0$. Following the same binning scheme as in Section~\ref{sec3.2.1}, the initial $e_0$ distribution is divided into 17 intervals ($e_{\rm bin}$) spanning $0.0-1.0$, with a interval width of 0.2 and a sliding step of 0.05. For each $e_0$ interval ($e_{\rm i}$), we define the conditional survival fraction as
\begin{equation}
f^* (t ; e_0 \in e_{\rm i}) =
\frac{N_b(t ; e_0 \in e_{\rm i})}
{N_b(t_0 ; e_0 \in e_{\rm i})}.
\end{equation}
Here, $e_0 \in e_\mathrm{i}$ indicates binaries whose initial eccentricity $e_0$ fall within the interval $e_\mathrm{i}$. $N_b(t ; e_0 \in e_{\rm i})$ is the number of surviving binaries at time $t$ that were initially assigned to the interval $e_\mathrm{i}$, while $N_b(t_0 ; e_0 \in e_{\rm i})$ is the corresponding initial number of binaries in that interval at $t_\mathrm{0}$. Binaries are removed from their $e_{\rm i}$ only when they are dynamically disrupted, thereby avoiding contamination caused by rapid eccentricity evolution and bin-to-bin migration.

\begin{figure*}[!t]
    \centering
        \includegraphics[width=1\columnwidth]{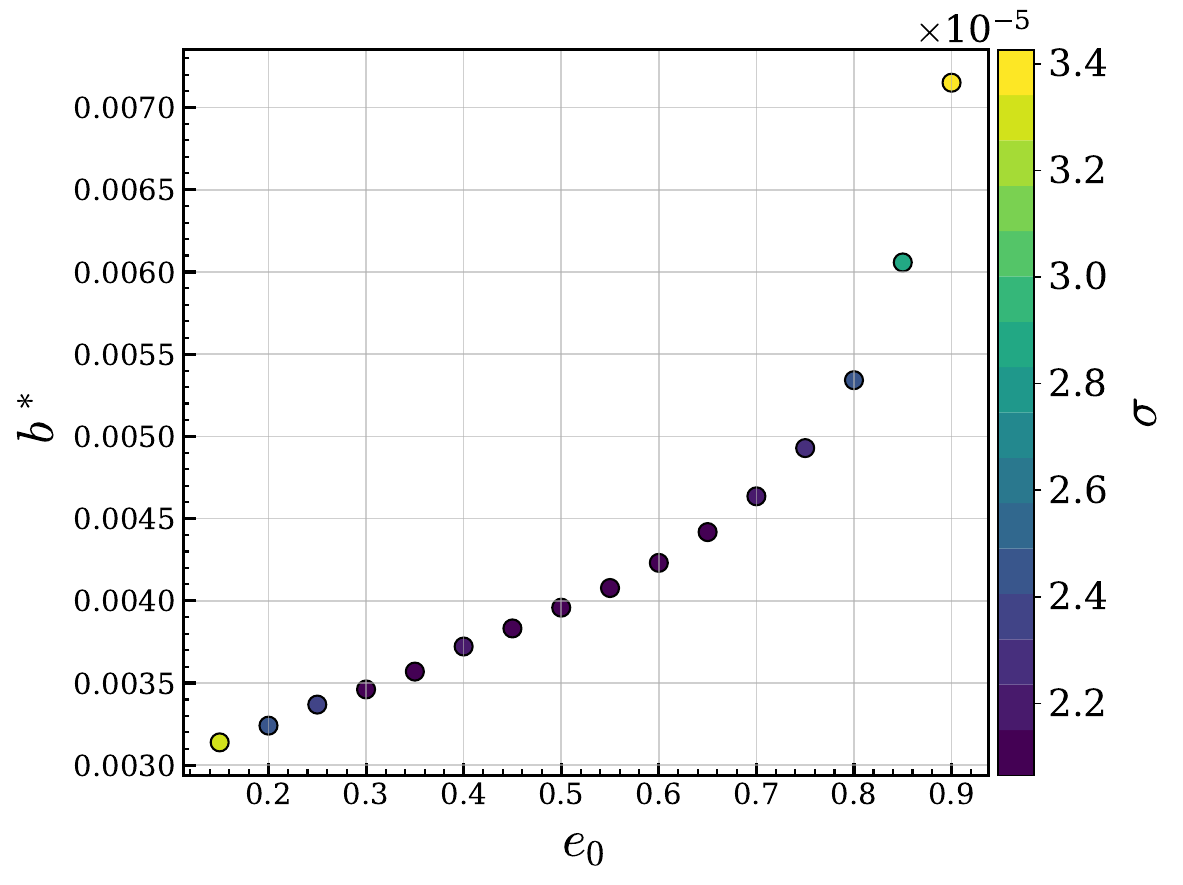}
    \hfill
        \includegraphics[width=1\columnwidth]{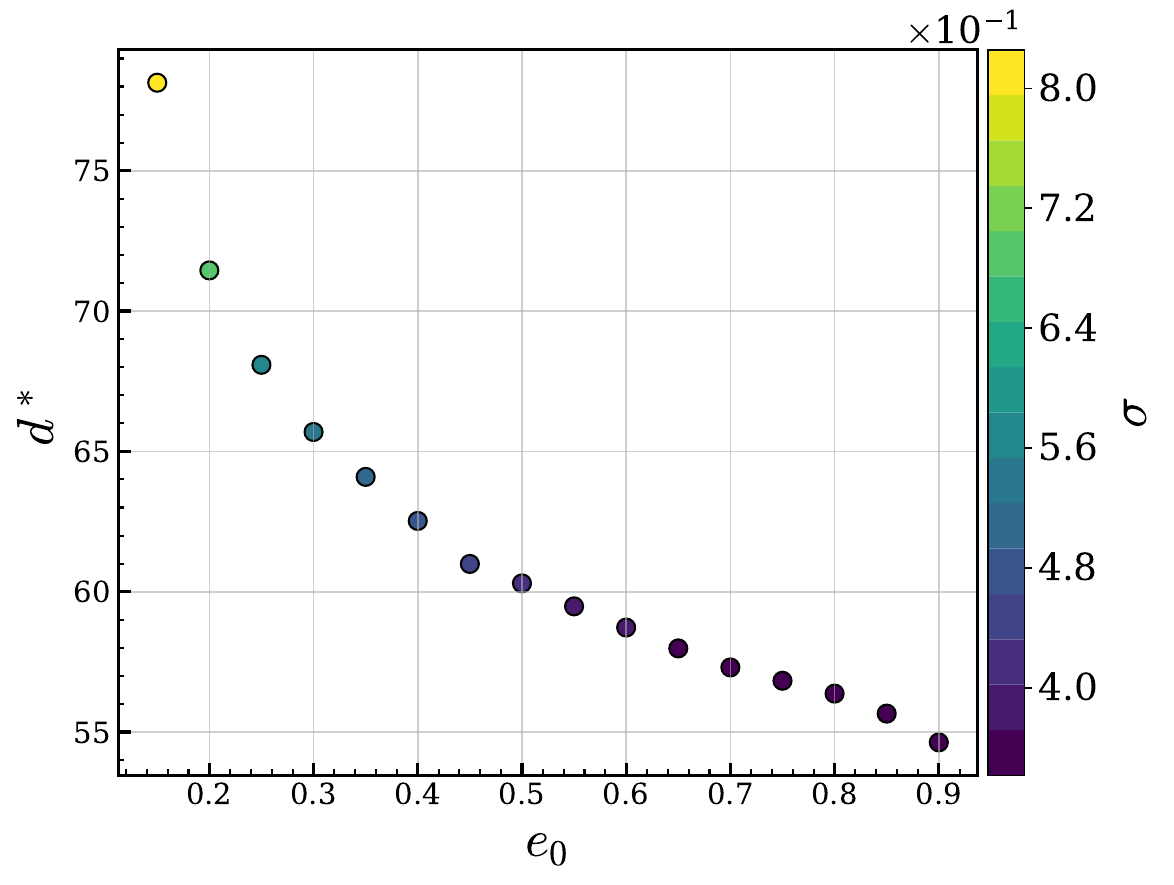}
    \caption{The relationship between the fit parameters $ \overline{k_1^*} $, $ \overline{t_\mathrm{b}^*} $, and $ e_0 $. The left subplot shows the relationship between the fitted parameter $ b^* $ and $ e_0 $ after fixing the parameter $ a^* $ (i.e., $0.5$) in Eq.~\ref{eq:all_k1_rho_fitting}. The right subplot shows the relationship between the fitted parameter $ d^* $ and $ e_0 $ after fixing the parameter $ c^* $ (i.e., $-0.46$) in Eq.~\ref{eq:tb_rho}. The horizontal axis of both subplots represents the median value of each $ e_0 $ interval, while the vertical axes represent the parameters $ b^* $ and $ d^* $, respectively. The color bars indicate the fit uncertainties for each parameter.}
    \label{fig:kb_td_ecc}
\end{figure*}

Under this definition, the evolution of each $e_{\rm i}$ can still be described by Eq.~\ref{eq:func} and obtain the corresponding $k_1$ and $t_b$, denoted as $k_1^*(e_0)$ and $t^*_{\mathrm{b}}(e_0)$. The dependence of mean parameters ( $\overline{k_1^*} (e_0)$ and $\overline{t^*_{\mathrm{b}}} (e_0)$ ) on the initial density, $\overline{k_1^*}(e_0,\rho_0)$ and $\overline{t^*_\mathrm{b}}(e_0,\rho_0)$, remains well described by the power-law relations in Eq.~\ref{eq:all_k1_rho_fitting} and Eq.~\ref{eq:tb_rho}. By fixing $a^*(e_0)$ in Eq.~\ref{eq:all_k1_rho_fitting} at 0.5  and $c^*(e_0)$ in Eq.~\ref{eq:tb_rho} at -0.46, as mentioned in Section~\ref{sec3.1}, we fitted the data to obtain the eccentricity-dependent parameters $b^*(e_0)$ and $d^*(e_0)$, shown in Figure~\ref{fig:kb_td_ecc}.

The lowest $e_0$ interval ($0.0-0.2$) is excluded from the fit due to its large dispersion. For the remaining intervals, $b^*(e_0)$ shows a clear positive correlation with $e_0$, indicating that binaries with larger $e_0$ are more susceptible to dynamical disruption at the same $\rho_0$. Conversely, $d^*(e_0)$ exhibits a inverse correlation with $e_0$, implying shorter disruption timescales for more eccentric systems under the same $\rho_0$.

\subsubsection{Dual Parameters: Mass Ratio and Orbital Period}\label{sec3.2.4}
Since binary disruption depends on both $q$ and $P$, these parameters may be correlated. 
We therefore analyze $f(t)$ in a two-dimensional parameter space of $\log_{10}(P)$ and $q$, combining their bins to form a grid in $ (\log_{10}(P), q) $. The $\log_{10}(P)$ bins remain unchanged with 21 intervals as in Section~\ref{sec:period}. Due to low binary counts per cell,  we redefined the $q$ bins with a width $ \Delta q = 0.5 $, starting from 0.0 and sliding by 0.05, resulting in 11 intervals. This yielded a total of 231 $ (\log_{10}(P), q) $ grid cells. We then tracked the temporal evolution of the binary population within each cell to characterize the dependence of $f(t)$ on the two-dimensional parameter space.

\begin{figure*}[!t]
    \centering
        \includegraphics[width=1\columnwidth]{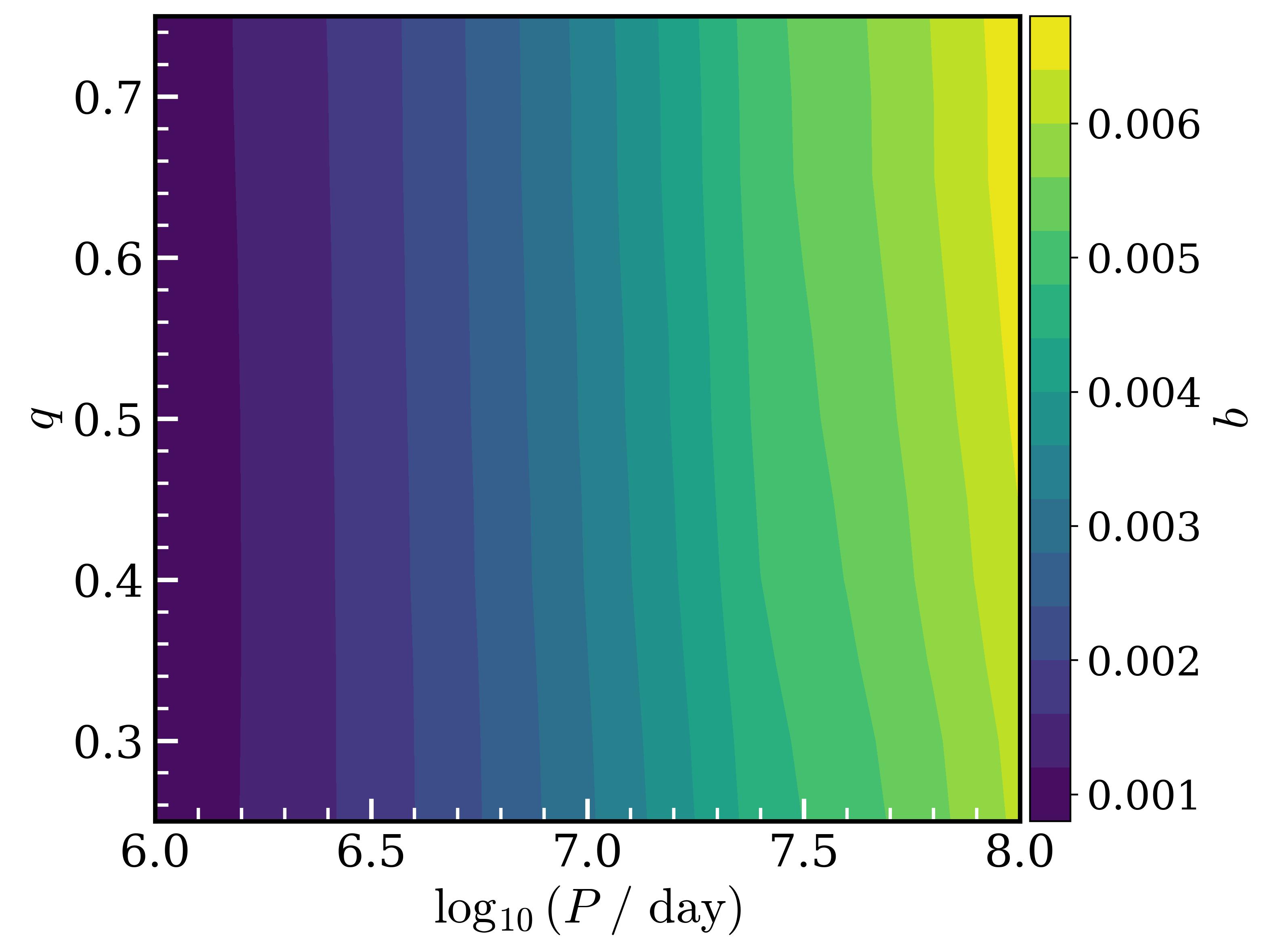}
    \hfill
        \includegraphics[width=1\columnwidth]{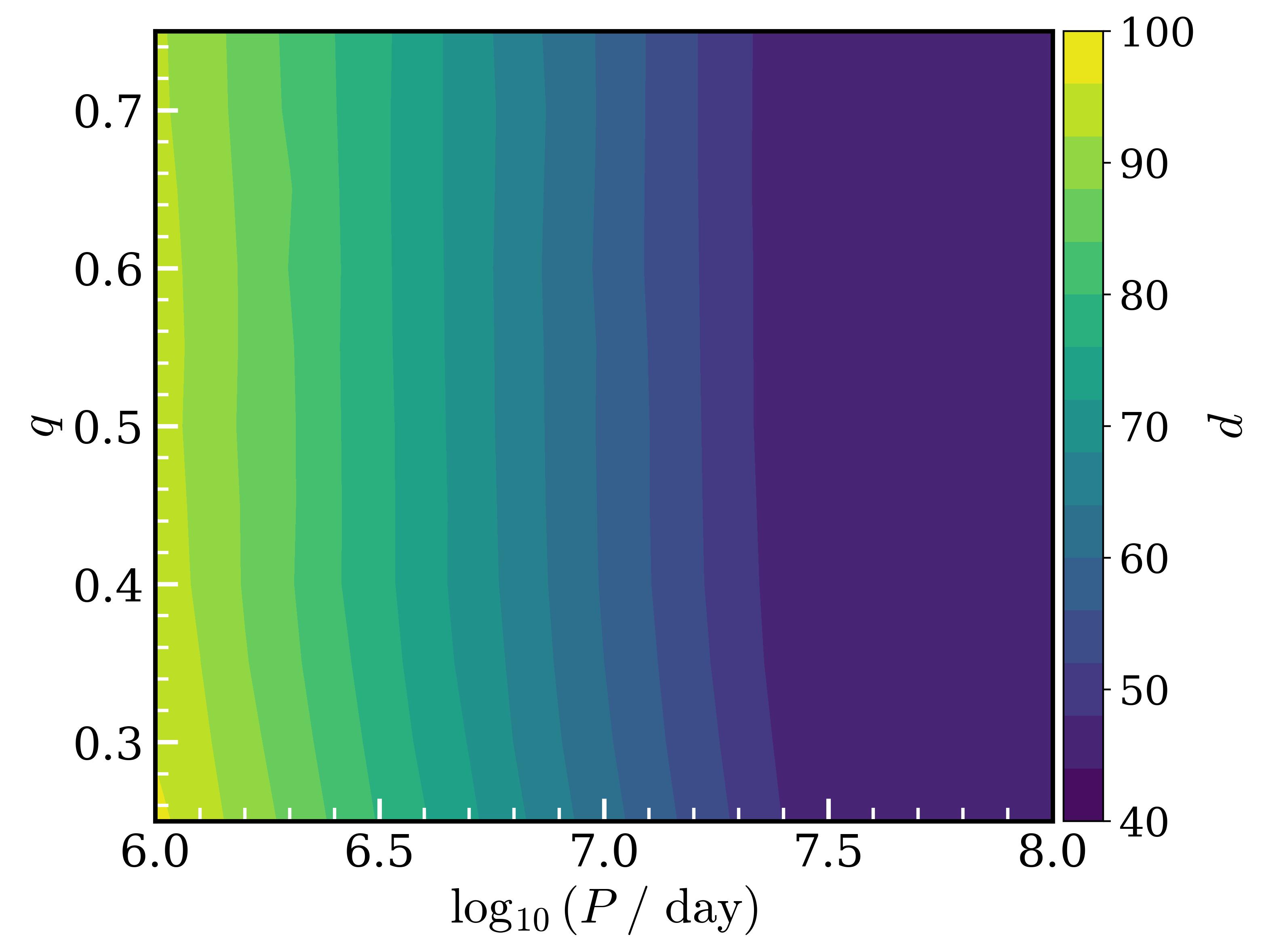}

    \caption{Relationships between the fitted parameters $b$ of $\overline{k_1}$ and $d$ of $\overline{t_\mathrm{b}}$ as functions of $P$ and $q$. The left panel corresponds to Eq.~\ref{eq:all_k1_rho_fitting} (fixed $a=0.62669 $), and the right to Eq.~\ref{eq:tb_rho} (fixed $c=-0.46$). Colors indicate the fitted parameter values $b$ and $d$ respectively, with axes showing the median values of each $(\log_{10}(P), q)$ grid. 
    }
    \label{fig:p_q_b_d}
\end{figure*}

Similar to the trends observed in Section \ref{sec3.2.1} and \ref{sec:period}, both $ \koneave$ and $\tbave$ versus $\rho_0$ in the $ (\log_{10}(P), q) $ grid follow a power law. 
Following the approach described in Section \ref{sec:period}, we allowed $a(P,q)$ to vary in each $(P,q)$ bin, obtaining values in the range 0.5979 to 0.6443. We adopt their mean value, $\langle a(P,q) \rangle = 0.62669$, as a fixed exponent when fitting $b(P,q)$, and similarly fix $c=-0.46$ when deriving $d(P,q)$. The results are shown in Figure~\ref{fig:p_q_b_d}.
The fit uncertainties of $b(P,q)$ increase with larger $b$, ranging from $0.00000468$ to $0.00002775$. 
In comparison, the uncertainties of $d(P,q)$ go from $0.29182$ to $0.94561$, which is larger due to the higher absolute values of $d(P,q)$.
Using these parameters, $\koneave(P,q,\rho_0)$ and $\tbave(P,q,\rho_0)$ can be determined for any given $\rho_0$, $q$, and $P$.

As shown in Figure~\ref{fig:p_q_b_d}, for binaries with longer orbital periods ($\log_{10}(P)\gtrsim 7.3$), the absolute value of $\koneave$ exhibits a positive correlation with both $P$ and $q$, indicating that systems with larger $P$ and $q$ tend to be disrupted more rapidly. In contrast, for shorter-period binaries ($\log_{10}(P)\lesssim 7.3$), the dependence of $|\koneave|$ on $q$ becomes weak.
In the right panel, we find that for $\log_{10}(P)\lesssim 7.3$ binaries, $\tbave$ shows an inverse correlation with period and mass ratio—binaries with smaller $P$ and $q$ require longer times to complete the disruption process.

\subsection{Comparison of Binary Fractions Inside and Outside the Tidal Radius}

Under Galactic tidal fields, stars and binaries escape from star clusters via tidal evaporation or ejection from close binary-single or binary-binary interactions.
Escaped (wide) binaries can survive long term, whereas inside clusters, close encounters and mergers disrupt binaries, and new binaries may form via gravitational capture \citep{Heggie2003}. These processes-dynamical interactions, tidal stripping, and mass segregation-produce substantial differences in the binary fraction between escapers and cluster members. 

To quantify this difference, we define the binary fraction as
\begin{equation}
    f_\mathrm{b} = \frac{N_\mathrm{b}}{N_\mathrm{s} + N_\mathrm{b}},
\end{equation}
where $N_\mathrm{b}$ and $N_\mathrm{s}$ denote the numbers of binary and single systems, respectively. We then analyze how the $f_\mathrm{b}$ differs inside and outside the tidal radius ($\rt$) at various stages of open cluster evolution. 
Here, $r_{\mathrm t}$ is adopted from the simulation. Then we can better understand clusters' contributions to the field binary population.

\begin{figure*}[!t]
    \centering
    \includegraphics[width=1\columnwidth]{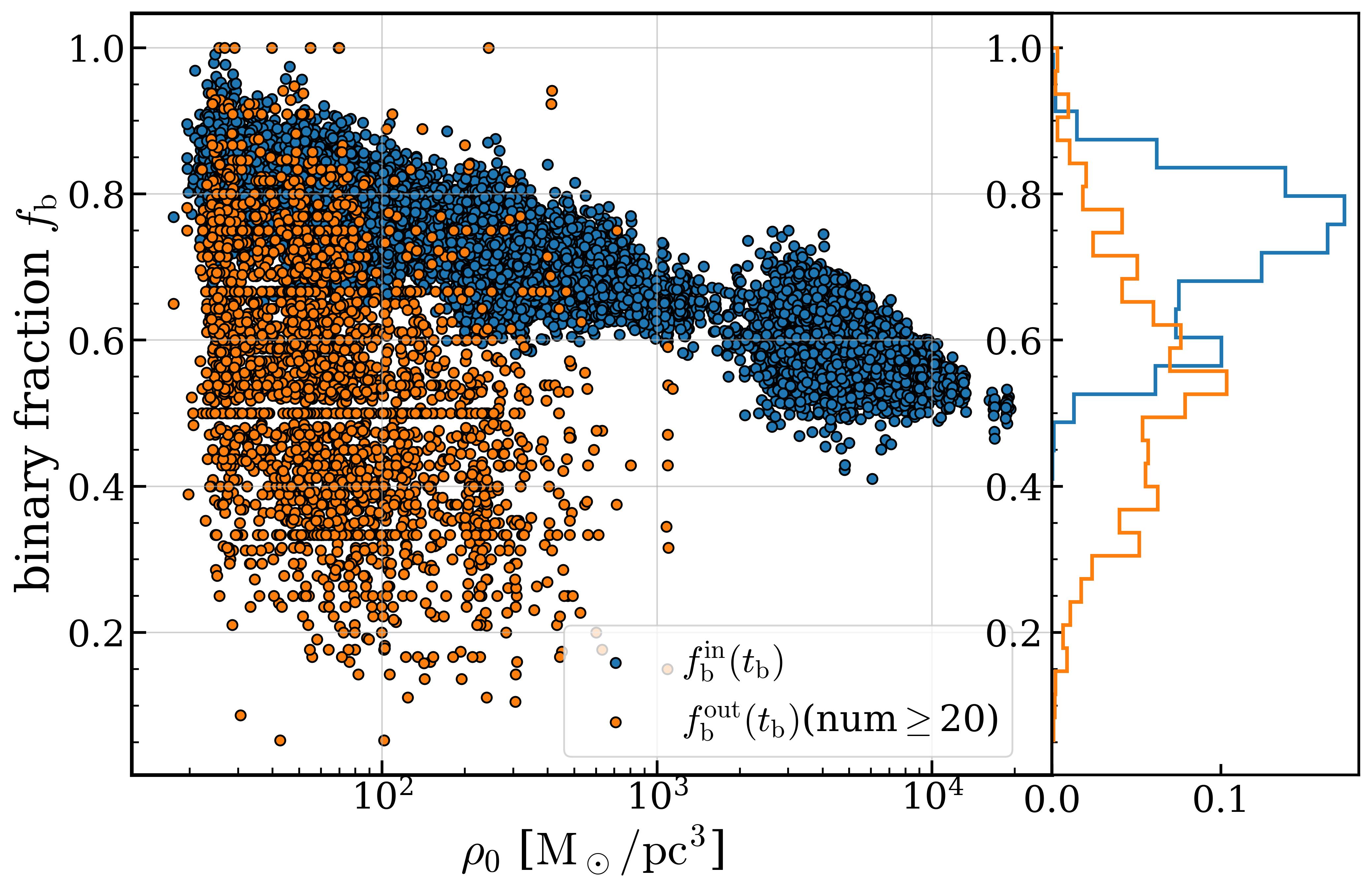}
    \label{x0_rtid_result}
    \hfill
    \includegraphics[width=1\columnwidth]{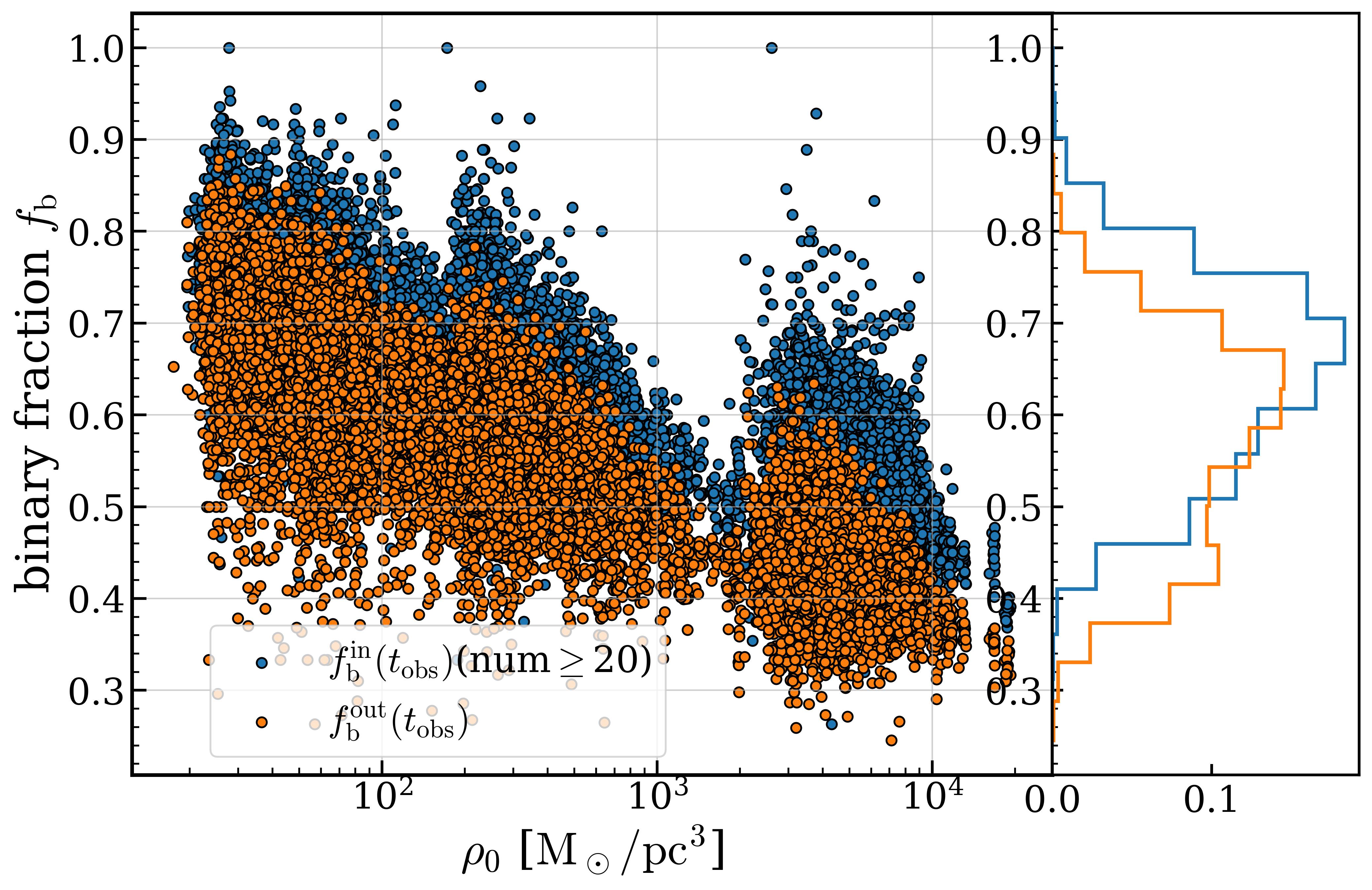}
    \label{age_rtid_result}
    
    \caption{Comparison of the binary fraction with initial cluster density ($\rho_\mathrm{0}$) at two evolutionary stages.
    Left: Binary fraction at the disruption timescale ($t_\mathrm{b}$). Blue points represent binaries within the tidal radius, and orange points represent those outside it. 
    The histogram shows the frequency distribution of the corresponding samples.
    Right: Same as the left panel, but for the observational timescale ($t_\mathrm{obs}$). 
    }
    \label{fig:x0_age_rtid_result}
\end{figure*}

We focus on comparing the changes in the $f_\mathrm{b}$ at two specific time points: $t_\mathrm{b}$ and the present time ($t_\mathrm{obs}$). At $t_\mathrm{b}$, the rapid binary-disruption phase ends.
As noted in Section \ref{sec3.1}, $ t_\mathrm{b} $ typically occurs before 20 Myr, during the early evolutionary stage of open clusters. 
At this time, wide binaries have undergone significant disruption, but the tidal stripping effect of the Galactic field on the cluster remains minimal, with most stars still gravitationally bound and only a small fraction ejected. 
The left panel of Figure \ref{fig:x0_age_rtid_result} illustrates 
$f_\mathrm{b}$ inside ($f^\mathrm{in}_{\mathrm{b}}$) and outside ($f^\mathrm{out}_{\mathrm{b}}$) the tidal radius as a function of $ \rho_{\mathrm{0}} $ at $t_\mathrm{b}$. 
As $ \rho_{\mathrm{0}} $ increases, 
$f^\mathrm{in}_\mathrm{b}(t_\mathrm{b})$ (blue points) gradually decreases due to enhanced disruption of wide binaries, consistent with the findings in Section \ref{sec3.1}.

For objects outside the tidal radius, some models have too few stars at $ t_\mathrm{b} $ to determine a meaningful $f^\mathrm{out}_\mathrm{b}(t_\mathrm{b})$. 
To address this, we require at least 20 stars to report statistics: data below this threshold are excluded from the figure.
Among statistically significant models  ($\ge 20$ stars), low $ \rho_{\mathrm{0}} $ yields a relatively dispersed  
$f^\mathrm{out}_\mathrm{b}(t_\mathrm{b})$ distribution, indicating high uncertainty in the number of escaped binaries, likely due to their much smaller population relative to those inside the clusters.
As $ \rho_{\mathrm{0}} $ increases, the number of models that meet the threshold ($\ge 20$ stars) declines rapidly and nearly vanishes above $10^3 M_\odot/\text{pc}^3$. This occurs partly because $ t_\mathrm{b} $ is earlier in high-density models, yielding fewer escapers, and partly because more massive clusters bind stars more strongly, further reducing escapers.

At time of observation  $t_\mathrm{obs}$, 
$f^\mathrm{in}_\mathrm{b}(t_\mathrm{obs})$ and $f^\mathrm{out}_\mathrm{b}(t_\mathrm{obs})$ 
represents the current binary fractions inside and outside clusters respectively.
The right panel of Figure \ref{fig:x0_age_rtid_result} illustrates how 
binary fraction 
depends on $ \rho_{\mathrm{0}}$ at $t_\mathrm{obs}$.
At $t_\mathrm{obs}$, clusters have typically evolved for longer than at $ t_\mathrm{b} $, leading to a larger number of escaped stars.
For low-$ \rho_{\mathrm{0}} $, the number of stars retained within the tidal radius can be so small that the inferred 
$f^\mathrm{in}_\mathrm{b}(t_\mathrm{obs})$ is statistically unreliable, whereas the population outside the tidal radius is generally large enough to avoid this issue.
Therefore, we no longer filter on the number of stars outside the tidal radius. Instead, we assess statistical significance within the tidal radius: for models with fewer than 20 stars inside the tidal radius, we deem  the binary  fraction insignificant and exclude them.

The right panel of Figure \ref{fig:x0_age_rtid_result} indicates that 
binary fraction 
outside the tidal radius is systematically lower than that inside, suggesting that binaries are more likely to remain bound within the cluster, while single stars preferentially escape. 
This phenomenon is likely driven by mass segregation, as the typically higher mass of binaries causes them to sink toward the cluster center and resist tidal stripping. 

\begin{figure*}[!t]
    \centering
    \includegraphics[width=1\columnwidth]{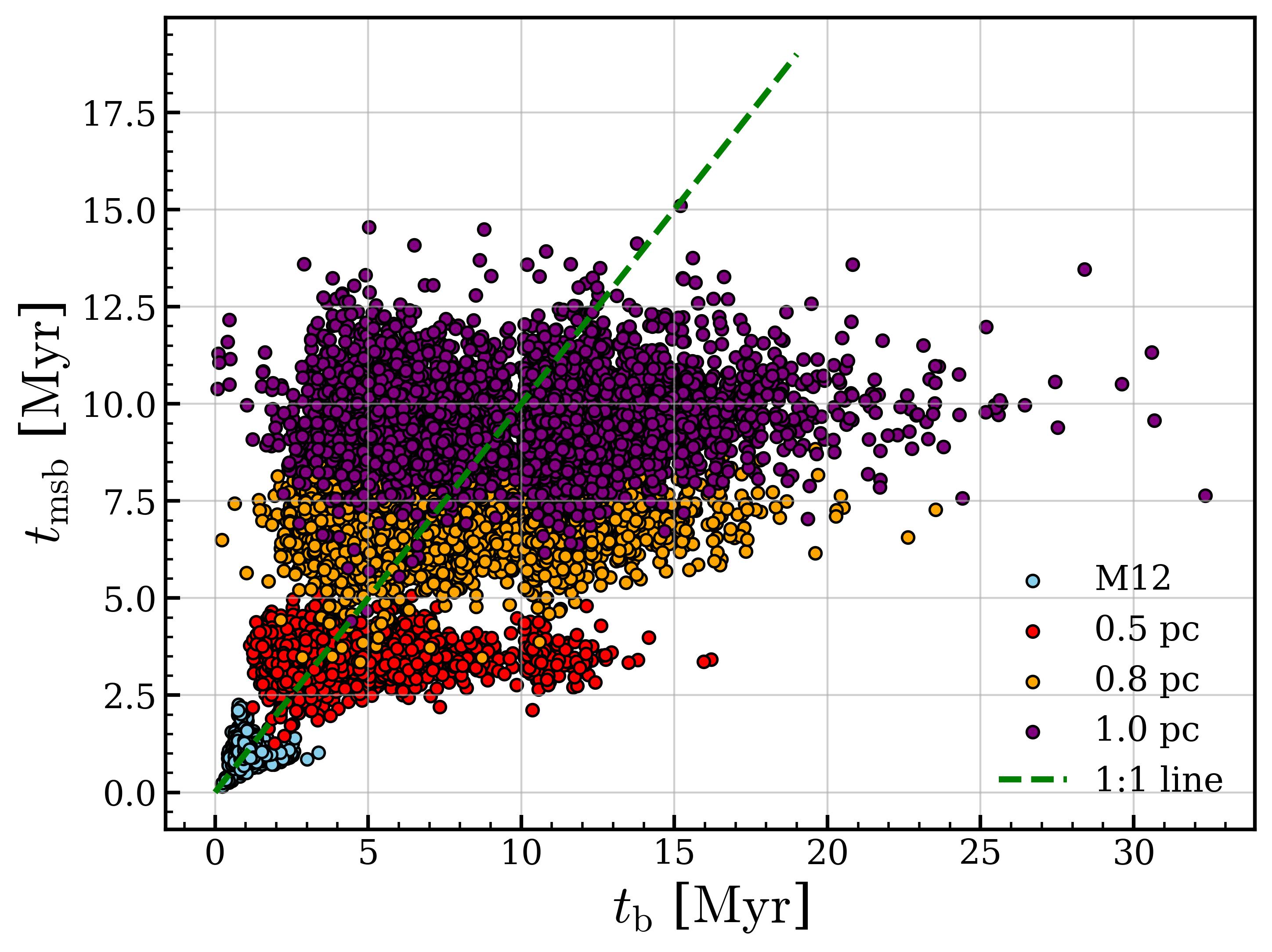}
    \label{tmsb_tb}
    \hfill
    \includegraphics[width=1\columnwidth]{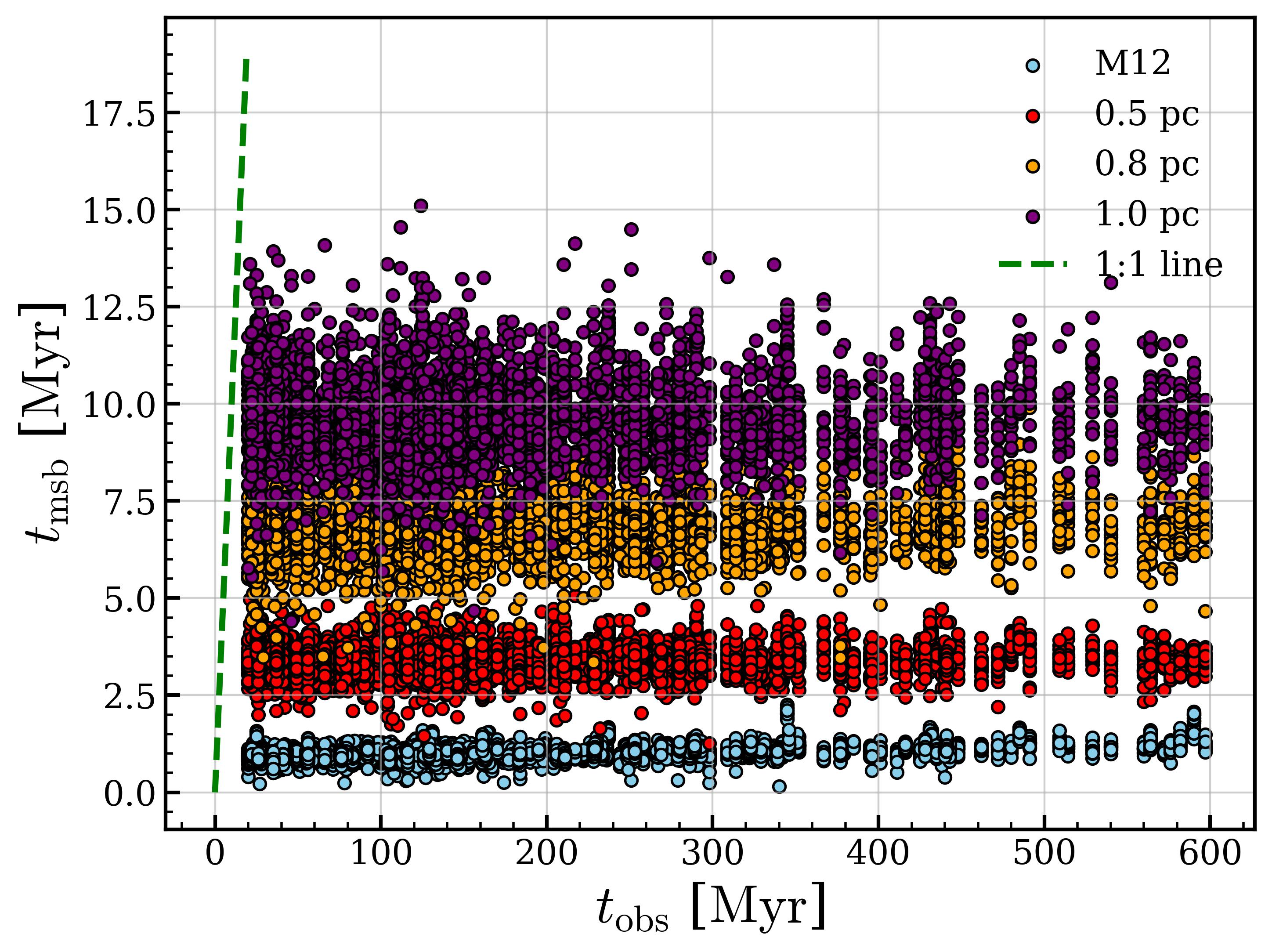}
    \label{msb_tobs}
    \caption{Comparison between the $t_\mathrm{b}$(left panel), $t_\mathrm{obs}$(right panel) and  mass segregation timescale of binaries ($\tmsb$) for all models. Blue, red, orange, and purple symbols correspond to models with $r_{\mathrm{h,0}}$ of \citetalias{64-Marks.2012}, 0.5, 0.8, and 1.0~pc, respectively. The green line indicates the one-to-one relation.}
    \label{fig:tmsb_tb_tobs}
\end{figure*}

According to \citet{84-Binney.1987}, the mass segregation timescale is given by
\begin{equation}
    \tms(m) \approx \frac{\langle m \rangle}{m} \trh,
\end{equation}
where $\langle m \rangle$ is the mean stellar mass and $\trh$ is the half-mass relaxation time, is related to $\tch$ by:
\begin{equation}
    \trh \approx \frac{0.1N}{\ln{N}}\tch.
\end{equation}
For equal-mass binaries with total mass $2\langle m\rangle$, $\tmsb \approx 0.5 \trh$.
Figures~\ref{fig:tmsb_tb_tobs} compare $t_\mathrm{b}$ and $t_\mathrm{obs}$ for all cluster models with the equal-mass binary segregation timescale $t_\mathrm{msb}$, which forms groups depending on  $r_{\mathrm{h,0}}$.
About half of  the models exhibit $t_\mathrm{b} > t_\mathrm{msb}$, suggesting that these clusters have already experienced at least one segregation timescale by $t_\mathrm{b}$. However, the limited number of escaped binaries makes this trend difficult to confirm in the left panel of Figure~\ref{fig:x0_age_rtid_result}. By $t_\mathrm{obs}$, all clusters satisfy $t_\mathrm{obs} \gg t_\mathrm{msb}$, indicating well-developed binary mass segregation, consistent with the right panel of Figure~\ref{fig:tmsb_tb_tobs}.



\begin{figure*}[!t]
    \centering
    \includegraphics[width=1.0\textwidth]{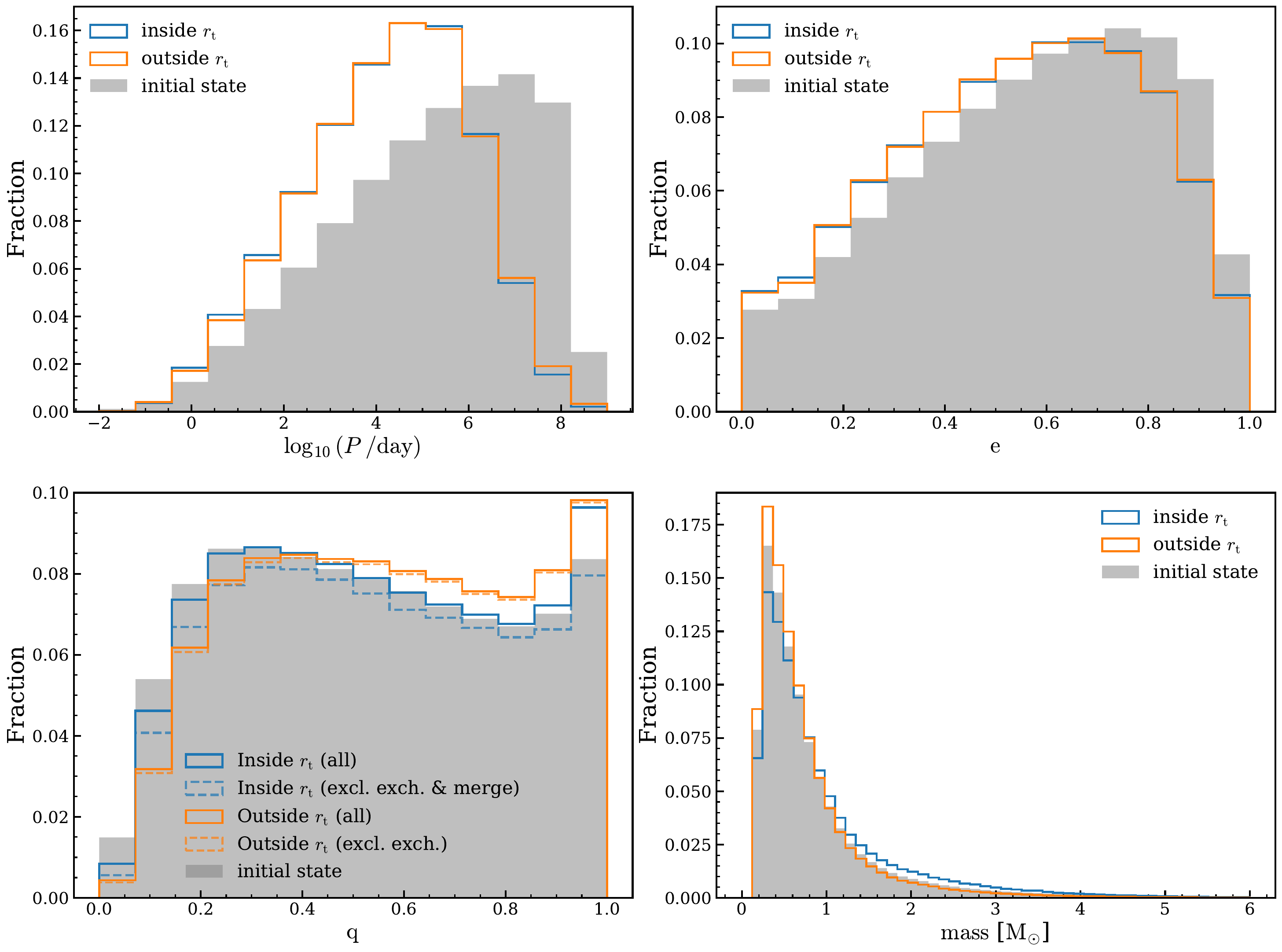}
    \caption{Comparison between distributions of orbital period (top left), eccentricity (top right), mass ratio (bottom left) and mass (bottom right) for binaries of all N-body models located inside (blue) and outside (orange) the tidal radius at $t_{\rm obs}$, along with the corresponding overall initial distributions (shaded). 
    The vertical axis represents the normalized fraction. The distributions at $t_{\rm obs}$ are normalized separately by the total number of binaries inside and outside the tidal radius, while the initial distributions are normalized by the total number of binaries at the beginning of the simulations. In the bottom left panel, dash lines represent the samples after excluding exchange and/or merger events. 
    }
    \label{tobs_hist}
\end{figure*}

Figure ~\ref{tobs_hist} illustrate the normalized distributions of $\log_{10}(P)$, $e$, $q$ and mass for binaries inside and outside $\rt$ at $\tobs$ from all models with initial half-mass radius $r_\mathrm{h,0}$ according to \citetalias{64-Marks.2012} , along with the corresponding initial distributions (shown as shaded regions).
Binaries inside and outside $\rt$ show nearly identical distributions in $P$ and $e$, 
while the $q$ distribution differs: binaries within $\rt$ tend to have smaller $q$. To examine whether this difference arises from dynamical exchanges or mergers, we excluded binaries that experienced such events and compared the remaining samples in the right panel of Figure~\ref{tobs_hist} (dashed lines). The results suggest that exchanges and mergers have only a minor effect and cannot account for the disparity in $q$ inside and outside clusters. The physical origin of this trend remains uncertain and needs further investigation.
Furthermore, binaries within $r_\mathrm{t}$ exhibit systematically higher masses than those outside, particularly for systems above $1\ \mathrm{M_\odot}$. The average binary mass inside $r_\mathrm{t}$ is about $0.99\ \mathrm{M_\odot}$, compared to $0.73\ \mathrm{M_\odot}$ outside. This result confirms a pronounced mass segregation effect within the cluster.

In addition, compared with the initial distribution, the binaries observed at later evolutionary stages exhibit shorter $P$ and lower $e$. 
This trend arises mainly from cluster dynamical interactions that preferentially disrupt wide binaries and thus lead to an increased fraction of short-period systems among the survivors.
Meanwhile, since high-$e$ binaries in the Kroupa binary model are mostly wide systems, their disruption further enhances the relative proportion of low-$e$ binaries. 
At $t_\mathrm{obs}$, except for binaries with $q \approx 1$, the distribution of relatively high-$q$ binaries ($0.5 \leq q < 1$) within $r_\mathrm{t}$ remains similar to the initial state. In contrast, outside $r_\mathrm{t}$, the fraction of high-$q$ binaries ($0.5 \leq q \leq 1$) increases relative to the initial distribution. The origin of this discrepancy needs further investigation.
Regarding the mass distribution, due to mass segregation, massive binaries are preferentially retained within $r_\mathrm{t}$, increasing the high-mass fraction relative to the initial distribution, while escaped binaries show a higher fraction at the low-mass end.

Finally, as shown in Figure \ref{fig:x0_age_rtid_result}, in the models with initial half-mass radius $r_\mathrm{h,0}$ according to \citetalias{64-Marks.2012} , the binary  fraction outside $r_\mathrm{t}$ is about $40$ to $50\%$, slightly higher than the approximately $30\%$ observed for field stars \citep{87-Liu.2019}, whose value is slightly underestimated due to the incompleteness of wide binaries (by $\sim13\%$). 
This discrepancy is likely due to the differences between the simulated and observed samples: our simulations assume all binaries are detectable, providing an upper limit. 
In contrast, observational constraints are limited by the spatial resolution, magnitude thresholds and detection methods. The method used by \cite{87-Liu.2019}, which relies on  magnitude differences between unresolved binaries and singles, mainly detects short-period binaries, resulting in a lower observed fraction. 


\subsection{Python tool for predicting binary survival fraction}

Previous analyses established an empirical relation between $f(t)$ and $P$, $q$, and $\rho_0$. 
Building on this, we developed an interpolation-based Python tool to predict binary survival fractions under various conditions:
\begin{enumerate}
    \item globally, as a function of $\rho_\mathrm{0}$ and time $t$ (Section \ref{sec3.1});
    \item as a function of $q$, $\rho_\mathrm{0}$, and $t$ (Section \ref{sec3.2.1});
    \item as a function of $\log_{10}(P)$, $\rho_\mathrm{0}$, and $t$ (Section \ref{sec:period});
    \item combining $q$, $\log_{10}(P)$, $\rho_\mathrm{0}$, and $t$ (Section \ref{sec3.2.4}).
\end{enumerate}
All tools are publicly available \footnote{Github link:  https://github.com/zhengzp3/oc-binary-disruption-tools.git}.

Because the slope $k_2$ in the second piecewise function in Equation \ref{eq:func} varies widely among $\rho_{0}$ and is generally small, our tools exclude binary disruption beyond $t_\mathrm{b}$. Thus, for $t > t_\mathrm{b}$, we set $k_2 = 0$ and assume the binary survival fraction remains constant.

\section{Discussion} \label{sec:4}

This work conducted extensive $N$-body simulations of OCs, but could not cover all possible initial conditions.
Firstly, we assumed all clusters initially follow a spherically symmetric Plummer density profile \citep{83-Plummer.1911}, a common but idealized model that differs from the irregular structures of real star-forming regions. In addition, observations indicate that young clusters are strongly affected by gas processes
, such as expulsion driven by stellar feedback \citep{89-Goodwin.2006,88-Baumgardt.2007,90-Farias.2018}. Future simulations will adopt an irregular density profile, such as fractal initial conditions with subclustering, and include gas-expulsion feedback to better understand their effects on binary evolution.

Second, the adopted primordial binary models \citep{29-Kroupa.1995} may not be fully consistent with all current observational constraints (e.g., \citealt{76-Moe.2017}; \citealt{10-Donada.2023}). Nevertheless, observational estimates of primordial binary properties remain subject to uncertainties arising from sample completeness and internal consistency, which make it challenging to construct a fully empirical initial model. The prescription of \cite{29-Kroupa.1995}, on the other hand, provides a unified theoretical framework for the primordial binary period distribution, allowing for systematic exploration of binary evolution across orbital scales. Furthermore, by comparing our $N$-body results with observed open clusters, we can assess the observational validity of this theoretical framework. In future simulations, we will implement observationally constrained binary properties to enable a more direct comparison with the present results.

\cite{60-Marks.2011} developed an analytic model of binary evolution across a wide range of initial density $\rho_0$ ($1-10^8 M_\odot/\text{pc}^3$). Our N-body models, based on observed OCs, cover the low-mass end (up to $10^4 M_\odot/\text{pc}^3$) and incorporate additional effects from stellar evolution and Galactic tidal disruption. The results generally agree with their predicted evolution of the binary energy distribution, although with significant statistical scatter. 

Third, the Galactic potential used in our simulations, \textsc{MWPotential2014}, is a simplified axisymmetric model comprising a bulge, disk, and dark matter halo, but it excludes complex structures such as the Galactic bar and spiral arms, which may influence the long-term tidal disruption of clusters. However, since internal dynamical interactions dominate the early disruption of wide binaries in open clusters over external tidal effects, the absence of these components is unlikely to significantly affect our main conclusions. In future studies, we will adopt more realistic time-dependent Galactic potentials, such as \cite{Hunter2024}, to better model tidal disruption and the distribution of escaped stars and binaries in the Galaxy.

Finally, our models do not necessarily reproduce the observed density profiles of OCs but only ensure a consistent total number of stars within the tidal radius. 
To resolve this limitation, future simulations will iteratively vary $r_\mathrm{h,0}$ together with $M_0$ to identify optimal initial conditions consistent with observations.

\section{Conclusion} \label{sec:5}
In this work, we primarily investigated the early binary disruption in open clusters. First, by filtering the 7,167 open clusters published in \cite{47-Hunt.2023}, we obtained 336 suitable cluster samples for study and performed simulations using the N-body simulation code \textsc{petar}. After completing the simulations, we plotted the evolution of the binary survival fraction over time for each N-body model. We found that the binary survival fraction evolution $f(t)$ evolves in two stages: a rapid initial decline followed by a slower decrease after a characteristic time $\tb$. Therefore, we used two piecewise linear functions to fit $f(t)$. The corresponding binary disruption rate $k_1$ during the rapid  phase follows a power-law dependence on $\rho_\mathrm{0}$ with significant scatter, yielding a power index of about $0.56$, close to $0.5$. This suggests that the early binary disruption rate scales with $\tch$. Further analysis indicates that this index is set by the close encounter timescale, influenced by gravitational focusing (Eq.~\ref{eq:coll_rate} and \ref{eq:coll_rate3}). 


The transition time $\tb$ follows a power-law relation with $\rho_0$, with a power index of about $-0.46$. We also found that $f(\tb)$ exhibits a weak power-law dependence on $\rho_\mathrm{0}$, suggesting a soft upper limit to binary disruption in OCs.


Next, we studied how early binary disruption depends on binary parameters, including mass ratio $q$, period $P$ and eccentricity $e$. Using a sliding window, we binned $q$ and $\log_{10}(P)$ individually and on a grid, computing $f(t)$ in each bin or cell. In all cases, $\koneave$ follows a power-law trend as the global case but the coefficient $b$ varies with $q$ and $P$.

In addition, to avoid bin-to-bin migration caused by rapid eccentricity evolution, we tracked binaries according to their initial eccentricity $e_0$. 
In this case，both $\overline{k_1^*}$ and $\overline{t^*_\mathrm{b}}$ follow the same power-law forms as in the Eq.~\ref{eq:all_k1_rho_fitting} and ~\ref{eq:tb_rho}, while the coefficients $b^*$ and $d^*$ vary with $e_0$. Specifically, $b^*$ shows a positive correlation with $e_0$, whereas $d^*$ exhibits a negative correlation. This implies that binaries with larger $e_0$ are more susceptible to dynamical disruption and have shorter disruption timescales under same $\rho_0$. 



After completing the analysis and quantifying the evolution of the binary survival fraction, we created a python tool for calculating binary survival fraction as a function of time, $q$, $P$, and $\rho_\mathrm{0}$.

We also investigated the binary fraction ($f_\mathrm{b}$) inside and outside the tidal radius of clusters at the end of the rapid binary disruption phase, $\tb$, and at the time of observation, $\tobs$. The results show that at both time points, the binary fraction within the cluster's tidal radius is higher, indicating that during the cluster's evolution, binaries tend to remain in the cluster's interior. This may result from mass segregation, where binaries sink toward the cluster center, and tidal stripping preferentially ejects single stars.  
Furthermore, by comparing the overall distributions of binaries at the initial and observational epochs, we find that the dynamical disruption of wide binaries leads to shorter orbital periods, lower eccentricities, and higher mass ratios in the surviving population.
In addition, the mass distribution analysis shows that binaries within the cluster tend to be more massive than those outside of it, further supporting the presence of mass segregation.


This study represents an initial step toward understanding the dynamical evolution of open clusters using Gaia data. Future work will incorporate more realistic initial conditions—including irregular morphology, gas expulsion, and time-dependent Galactic potentials—to better understand binary evolution, the initial states of observed clusters, influence of black holes, and their contribution to field stars and binaries.

\begin{acknowledgments}
    L.W. thanks the support from the National Natural Science Foundation of China through grant 12573041, 12233013 and 21BAA00619, the High-level Youth Talent Project (Provincial Financial Allocation) through the grant 2023HYSPT0706, the one-hundred-talent project of Sun Yat-sen University, the Fundamental Research Funds for the Central Universities, Sun Yat-sen University (2025QNPY04).
    The authors acknowledge the Beijing Beilong Super Cloud Computing Co., Ltd for providing HPC resources that have contributed to the research results reported within this paper.URL: http://www.blsc.cn/.
\end{acknowledgments}

\software{numpy (\citealp{85-Harris.2020}),
          matplotlib (\citealp{86-Hunter.2007}),
          astropy (\citealp{astropy2013..2013A&A...558A..33A,astropy2018..2018AJ....156..123A,astropy2022..2022ApJ...935..167A}),
          \textsc{sdar} (\citealp{68-Wang.2020}, https://github.com/lwang-astro/SDAR),
          \textsc{petar} (\citealp{66-Wang.2020}, https://github.com/lwang-astro/PeTar),
          \textsc{fdps} (\citealp{69-Iwasawa.2016,70-Iwasawa.2020,71-Namekata.2018}, https://github.com/FDPS/FDPS),
          \textsc{sse/bse} (\citealp{72-Hurley.2000,73-Hurley.2002,74-Banerjee.2020}),
          \textsc{galpy} (\citealp{75-Bovy.2015}, modified version: https://github.com/lwang-astro/galpy),
          \textsc{mcluster} (\citealp{65-Kupper.2011,Wang2019}, modified version: https://github.com/lwang-astro/mcluster),
          \textsc{galevnb} (\citealp{63-pang.2016}),
          \textsc{pygaia} (the Gaia Project Scientist Support Team and the Gaia Data Processing and Analysis Consortium (DPAC), https://github.com/agabrown/PyGaia.git)
          }

\appendix

\bibliography{paper}{}
\bibliographystyle{aasjournalv7}

\end{CJK*}
\end{document}